\documentclass[letterpaper,twocolumn,10pt]{article}
\usepackage{usenix,epsfig,endnotes}

\usepackage{hyperref}
\usepackage{color}

\usepackage[english]{babel}
\usepackage[utf8]{inputenc}
\usepackage{amsmath}
\usepackage{amsfonts}
\usepackage{graphicx}
\usepackage[colorinlistoftodos]{todonotes}
\usepackage{algorithm}
\usepackage{algpseudocode}
\usepackage{float}

\usepackage{authblk}

\begin{document}

\date{}

\title{\Large \bf Detecting Cyber-Physical Attacks in Additive Manufacturing \\ using Digital Audio Signing}

\author[1]{Sofia Belikovetsky}
\author[1]{Yosef Solewicz}
\author[2]{Mark Yampolskiy}
\author[3]{Jinghui Toh}
\author[1,3,4]{Yuval Elovici}
\affil[1]{Ben-Gurion University of the Negev}
\affil[2]{University of South Alabama}
\affil[3]{Singapore University of Technology and Design}
\affil[4]{Cyber Security Research Center of the Ben-Gurion University (CSRC)}
%


\maketitle

\thispagestyle{empty}

\subsection*{Abstract}

Additive Manufacturing (AM, or 3D printing) is a novel manufacturing technology that is being adopted in industrial and consumer settings.
However, the reliance of this technology on computerization has raised various security concerns. 
In this paper we address sabotage via tampering with the 3D printing process.
We present an object verification system using side-channel emanations: sound generated by onboard stepper motors.
The contributions of this paper are following.
We present two algorithms: one which generates a master audio fingerprint for the unmodified printing process, and one which computes the similarity between other print recordings and the master audio fingerprint. 
We then evaluate the deviation due to tampering, focusing on the detection of minimal tampering primitives. 
By detecting the deviation at the time of its occurrence, we can stop the printing process for compromised objects, thus save time and prevent material waste.
We discuss impacts on the method by aspects like background noise, or different audio recorder positions.
We further outline our vision with use cases incorporating our approach. 

\section{Introduction}

Additive Manufacturing (AM), often 
referred to as 3D printing, is a manufacturing 
technology that creates parts and prototypes by incrementally fusing layers of material together. 
This manufacturing technology can create objects from polymers, metals and alloys, and composites. 

AM 
has numerous technological, environmental and socioeconomic advantages. 
These include the ability to manufacture objects with complex internal structures, 
shorter design-to-production time, just-in-time and on-demand production, and reduce source material waste. 
These advantages enable a broad range of applications, ranging from models and prototypes up to functional parts in safety-critical systems. 
A recent example of the latter is the FAA-approved 3D-printed fuel nozzle for the GE's state of the art LEAP jet engine~\cite{GE2015faa}.


According to the \emph{Wohlers Report}, a renowned annual survey of advances in AM, in 2015 the AM industry accounted for \$5.165 billion of revenue, with 32.5\% of all AM-manufactured objects used as functional parts. 
A study conducted by \emph{Ernst \& Young}~\cite{ey2016how} 
shows rapidly growing adoption of this technology worldwide. In the U.S. alone, 16\% of surveyed companies have experience with AM and another 16\% are considering adopting this technology in the future. 

Due to the growing importance of AM and its reliance on computerization, many researchers have raised security concerns. 
So far, two major threat categories have been identified for AM: (1) sabotage~\cite{sturm2014cyber, yampolskiy2014towards, yampolskiy2015security, zeltmann2016manufacturing, yampolskiy20163dpaaw, belikovetsky2017dr0wned, moore2017implications} and (2) violation of Intellectual Property (IP)~\cite{yampolskiy2014towards, yampolskiy2014intellectual, brown2016legal, faruque2016acoustic}. Sabotage attacks aim to inflict physical damage, e.g., by compromising part quality or by damaging AM equipment. IP violation attacks aim to illegally replicate 3D objects or the manufacturing process itself. Additionally, several articles discuss using 3D printers to manufacture illegal items, e.g., firearms, or components of explosive devices~\cite{blackman20141st, mcmullen2014worlds, johnson2013print}.

\begin{figure}[tbp]
	\centering
		\includegraphics[width=.45\textwidth]{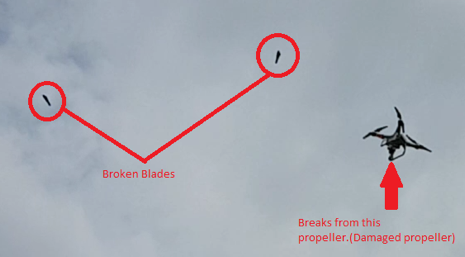}
	\caption{Sabotaged Quadcopter's Propeller breaks during Flight (\emph{dr0wned} study~\cite{belikovetsky2017dr0wned})}
	\label{fig:drone-in-flight-breaks}
\end{figure}

This paper exclusively focuses on sabotage attacks and proposes a method of detecting such attacks.
The importance of combating these attacks is illustrated by the recent \emph{dr0wned} study~\cite{belikovetsky2017dr0wned}.

In the study, the researchers presented a full chain of attack with AM, and introduced a novel cyber-physical attack impacting the material fatigue of a functional part. The authors sabotaged the 3D printed propeller of a quadcopter UAV, causing the propeller to break and the quadcopter to fall from the sky after a short period of flight time  (see Figure~\ref{fig:drone-in-flight-breaks}). 
While this only led to the loss of a \$1000 drone, similar attacks on functional parts for safety-critical systems may cause tremendous monetary losses, disruptions, and loss of human life.

In this paper, we propose a method capable of detecting 
such cyber-physical attacks via side-channel audio verification. 

The remainder of the paper is structured as follows: we first present the considered threat model on the manufacturing process in section \ref{sec:ThreatModel}. Next we introduce the proposed solution in section \ref{sec:ProposedSolution}. Section \ref{sec:ExperimentalEvaluation} presents an evaluation of the detection capabilities of the solution and its limitations. In section \ref{sec:use_cases}, we discuss various use cases that can deploy the proposed solution. 
Lastly, we discuss previous work in this area in section \ref{sec:related_work}.

\section{Considered Threat Model}
\label{sec:ThreatModel}

%
Although the central function of any AM process is the 3D printing itself, there are many stages involving other equipment and technologies.
Typically, a 3D object's blueprint (in STL, AMF, or 3MF file format) is first stored on a computer. 
Before printing, the 3D model is ``sliced'' by a program into individual layers.
For desktop 3D printers that employ \emph{Fused Deposition Modeling} (FDM) technology, the open source softwares like \emph{Slic3r} or \emph{Cura} are commonly used. The parameters like ``fill density'' and ``fill pattern'' influence how the source material is actually deposited in an individual layer.
The description of these layers can vary greatly between different AM technologies.

The tool path generated by this stage is then transmitted to 3D printer via USB, SD card or network connection.
The tool path is commonly composed of \emph{G-Code} commands, a legacy language for CNC machines.
The individual G-Code commands are interpreted by the firmware installed on a 3D printer and translated to electrical signals for individual actuators, such as motors for X/Y/Z movement and filament extrusion, or a heater nozzle, etc.

In this workflow, cyber threats arise because each of the 3D model representations can be corrupted. 
Researchers have shown that the original blueprint file can be corrupted via remote access to the computer~\cite{belikovetsky2017dr0wned} or by malware running on a computer~\cite{sturm2014cyber}. 
Vulnerabilities in network communications can be exploited to alter print jobs~\cite{do2016data}.
It has been suggested that models can be modified or even completely substituted by malicious 3D printer firmware~\cite{moore2017implications}.

Regardless of the representation compromised, 
the Cyber-\emph{Physical} impact depends on the physical change of the printed object.
Researchers have shown that changes to 3D model alone\footnote{Changes of manufacturing process as discussed in~\cite{yampolskiy2015security} are out of scope of this paper.} can
prevent its integrability~\cite{xiao2013security, moore2017implications},
reduce its tensile strength~\cite{sturm2014cyber, zeltmann2016manufacturing},
or impact its fatigue life~\cite{belikovetsky2017dr0wned}.
Especially in the case of a functional part, such changes can lead to the destruction of a CPS employing this part, as shown in the recent \emph{dr0wned} study~\cite{belikovetsky2017dr0wned}.

\section{Proposed Solution}
\label{sec:ProposedSolution}

In this section, we propose a solution for the detection of the sabotage attacks that change a 3D printed object's geometry.
We describe how the 3D printing process' audio fingerprints can be generated and verified.
We conclude this section with the description of how two audio fingerprints can be compared.

\subsection {General Concept: Verification via Audio Side-Channel Fingerprinting}

While the protection of every translation stage and representation of a 3D object description is theoretically possible, it has numerous drawbacks.
From the operations point of view, this would require the protection of multiple stages with negative impacts on the overall performance of the 3D printing process.
In the worst case, introducing security measures could interfere with time-critical processes, degrading a manufactured part's mechanical properties.
From the security stand point, the complexity of such a solution will likely be accompanied by new vulnerabilities. Furthermore, if the security mechanisms are integrated into equipment involved in the 3D printing process, any malicious code that can change the process can also disable or bypass the security mechanisms.
Therefore, there is need for a verification method that is independent of the manufacturing process equipment. 

\begin{figure}[htbp]
	\centering
	\includegraphics[width=.45\textwidth]{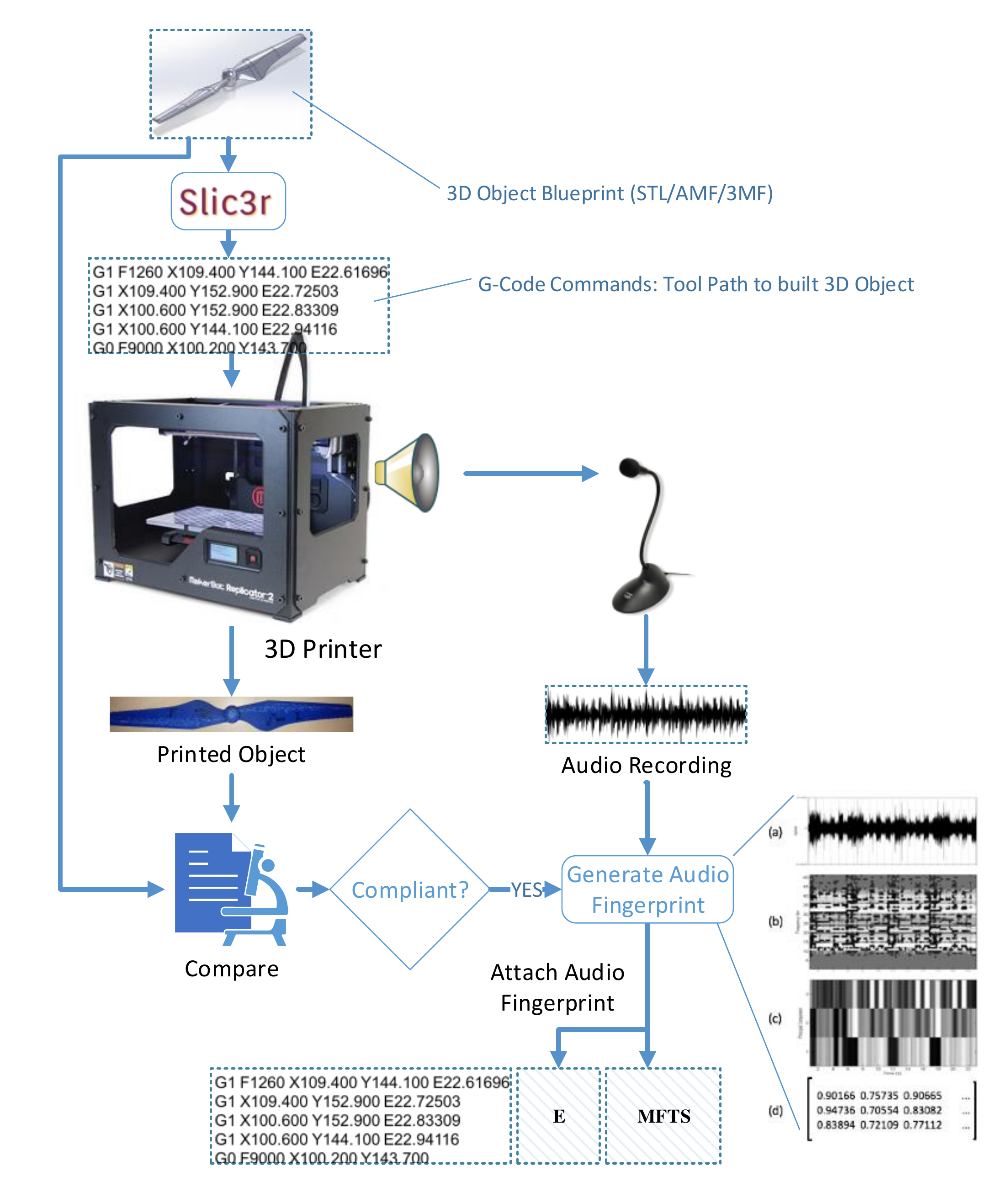}
	\caption{Audio Fingerprint Generation}
	\label{fig:AudioFingerprintGeneration}
\end{figure}

%
%
Similar to the KCAD approach of Chhetri et al.~\cite{chhetri2016kcad}, we exploit that in the FDM technology, the geometry of a printed object is defined by the movements of four stepper motors (for X/Y/Z axes and filament extrusion), each of which generate noise with unique characteristics. 
To detect manipulations, we propose to record and digitally sign the sound generated from manufacturing a verifiable benign 3D object. 
As with an approach proposed for the detection of hardware Trojans~\cite{agrawal2007trojan}, after the sound is recorded, the compliance of the printed 3D object to the blueprint can be validated using destructive methods; only then the recorded sound can be used as a ``fingerprint'' of a valid manufacturing process. The workflow of this process is illustrated in figure \ref{fig:AudioFingerprintGeneration}. The input of this workflow is either an STL or G-code file of an object that is produced in a lab environment and the audio signal is recorded. The fingerprint on the audio signal is calculated (subsection \ref{subsec:fingerprintGeneration}), encrypted, and concatenated to the G-code file. 
When the same 3D object is manufactured again, its validity can be verified by comparing the sound generated during the manufacturing process to the sound of the signed fingerprint. 

In industrial settings, when manufacturing large runs of the same 3D object, fingerprint generation can be performed and verified by the manufacturer.

\subsection{Master Audio Fingerprint Generation}
\label{subsec:fingerprintGeneration}

There are several approaches for audio fingerprinting, depending on the tasks and challenges involved~\cite{cano2005review}.
The scheme used for this paper is inspired by the idiosyncrasies of the noise emitted by the mechanical components of 3D printers. 
As shown in~\cite{faruque2016acoustic}, all four stepper motors on an FDM 3D printer produce noise with unique characteristics; furthermore, these characteristics distinguish a motor movement's direction and, loosely, speed.
Therefore, we claim that similar motor movements will lead to similar acoustic patterns that can be parametrized and matched to ensure manufacturing process authenticity.

3D printing acoustic patterns are limited to roughly fixed patterns concentrated in specific frequency ranges, since they are generated by a fixed combination of mechanical transitions.
A common audio fingerprinting approach is to create a summary of an audio recording by parametrizing unique acoustic anchor points in frequency and time. 
\begin{figure}[tbp]
	\centering
	\includegraphics[width=.4\textwidth]{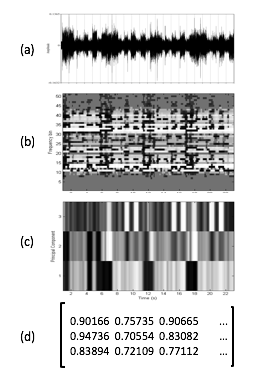}
	\caption{Audio Fingerprint Generation: (a) original signal, (b) spectrum after running FFT, (c) gray-scale representation after dimension reduction with PCA algorithm, (d) numeric representation of audio fingerprint}
	\label{fig:alg_w_pics}
\end{figure}

We propose to generate a fingerprint of a FDM 3D printing process via the following steps (see Figure~\ref{fig:alg_w_pics}).
First, we divide the original audio recording into equidistant overlapping time frames and apply a Fast Fourier Transform (FFT) on each of these frames.
We then use Principal Component Analysis (PCA)~\cite{pearson1901liii}, which compresses the data by reducing the number of dimensions with minimal loss of information. The use of PCA enables pattern identification in the signal and the comparison of different signals.
The outcome of the PCA transformation step can then be represented as a matrix.

The output of this approach 
is a text file containing the audio fingerprint. 
This can be used as a master file textual summary (MFTS) for the printing process.

\begin{algorithm}
	\begin{algorithmic}[1]
		\Function{AudioFingerprintCreation}{$signal$}

		\State $TrimByMarkers(signal)$ \label{alg1:trim}
		
		\State $downsample = Resample(signal, 2000)$ \label{alg1:resampling}
		\State $S = spectrogram(downsample, 0.75,0.1,1000,20)$ \label{alg1:FFT}
		\State $S = S - mean(S)$ \label{alg1:mean}
		\State $covariance = S*S^T$ \label{alg1:cov}
		\State $[E] = eigs(covariance, 3)$ \label{alg1:eigs}
		\State $E = E/norm $ \label{alg1:norm}

		\State $MFTS=S*E$ \label{alg1:project}
		
		\Return $<E, MFTS>$
	
		\EndFunction
		
	\end{algorithmic}
		\caption{Audio Fingerprint Generation}
		\label{alg:createMFTS}
\end{algorithm}

Algorithm~\ref{alg:createMFTS} 
depicts the pseudo code of the algorithm we use to generate the master fingerprint.

\begin{description}
	\item[Line~\ref{alg1:trim}:]  
	First, we bound the audio signal to the section directly related to the manufacturing process. 
	To synchronize between the audio recording device and the 3D printer, we insert audible markers at the beginning and end of the 3D printing process.
	We used the beep command (M300) and the Dwell command (G4) to signal the boundaries. 
	This allows us to trim off irrelevant data.

	\item[Line~\ref{alg1:resampling}:]  
	Preliminary experiments indicated that, for our 3D printer, a bandwidth of 1kHz captures most of the relevant acoustic information. 
	Therefore, according to the Nyquist rate, the original audio recording can be downsampled to 2kHz without introducing errors. 
	The dowsampling step includes a low-pass filtering of all signals with frequency above 1kHz. 
	Downsampling reduces the computation difficulty of further steps and discards less-informative high frequency regions.
	
	\item[Line~\ref{alg1:FFT}:]  
	Then the spectogram showing the power density of the downsampled audio record is calculated\footnote{A spectrogram can be created by sequentially calculating the magnitude of the spectrum of overlapping frames of the signal using a Fast Fourier Transform (FFT) implementation like~\cite{frigo1998fftw}.}. 
	We have selected the following spectrogram parameters: the signal is segmented into overlapping frames of 0.75 seconds with a stepping factor of 0.1 second; the FFT resolution is 20 Hz, resulting in 50 bins up to 1000 Hz, the signal bandwith. 
	The spectrum of each frame generates a gray-level column along the frequency axis at the corresponding signal time slot. Darker levels represent higher energy densities and brighter levels represent lower energy densities (image (b) in Figure~\ref{fig:alg_w_pics}).

	\item[Line~\ref{alg1:mean} through \ref{alg1:norm}:]  
	Next, we apply Principal Component Analysis (PCA)~\cite{pearson1901liii}, a technique that compresses the data by reducing the number of dimensions, without much loss of information. 
	%
	The PCA transformation consists of several steps. 
	First,  the data is centered by removing the mean spectrum (static component) from each frequency bin in step \ref{alg1:mean}. This helps remove potential channel mismatch between the current recording and future recordings. 
	Then the data covariance is calculated (line ~\ref{alg1:cov}). Covariance measures the ``spread'' of a set of points around their center of mass (mean). Thus, we measure how much the dimensions vary from the mean with respect to each other. 
	We then calculate the eigenvectors of the covariance matrix in step~\ref{alg1:eigs} and normalize them in step~\ref{alg1:norm}.
	%

	The eigenvalues in PCA indicate the data variance that is associated to a specific eigenvector. Therefore, the highest eigenvalue indicates the highest variance in the data was observed in the direction of its eigenvector. Accordingly, by using all the eigenvectors, we can represent all the variance in the data. In figure ~\ref{fig:eigenvectors}, we show a graph of the explained variance gained by adding each extra eigenvector. PCA, besides compression, may help reduce noise by eliminating secondary effects found on the less significant eigenvectors.
	%
	We have empirically identified 
	that 3 eigenvectors are sufficient to represent the recordings for audio fingerprint generation and comparison. 

\begin{figure}[tbp]
	\centering
	\includegraphics[width=.45\textwidth]{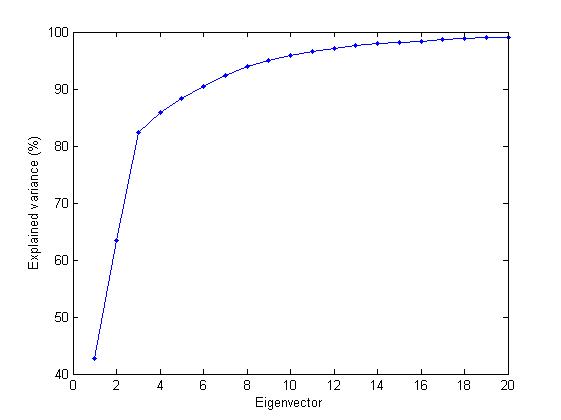}
	\caption{Eigenvalues of the calculated eigenvectors}
	\label{fig:eigenvectors}
\end{figure}

	\item[Line~\ref{alg1:project}:]  
	PCA uses the main eigenvectors of the covariance of the observed data to project it onto an orthogonal low-dimensional subspace. The learned subspace is shown to be closely related to the subspace spanned by the data centroids obtained through unsupervised clustering \cite{ding2004k}. These centroids are expected to summarize the set of acoustical patterns corresponding to the printer set of actions. 
	Therefore, the final step (~\ref{alg1:project}) of the algorithm is to project the spectogram matrix onto the three selected eigenvectors, which generate a stream of vectors of length three every 0.1 second.

	\item[Algorithm output:]  
	The outputs of the algorithm are the MFTS and the three selected eigenvectors that were calculated for the audio master file.

\end{description}


%


\subsection{Audio Fingerprint Comparison}

\begin{figure}
	\centering
	\includegraphics[width=.45\textwidth]{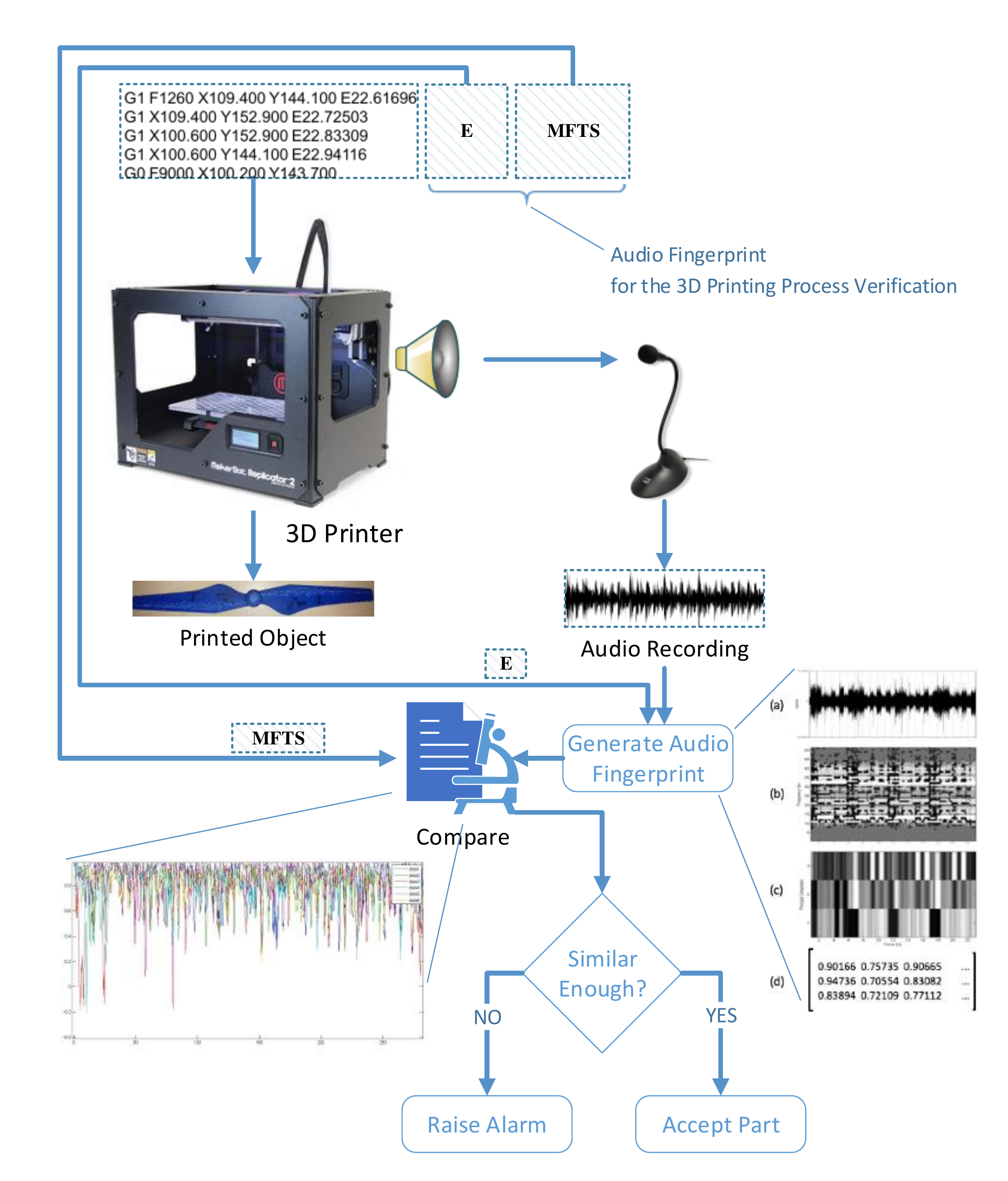}
	\caption{Audio Fingerprint Verification}
	\label{fig:AudioFingerprintVerification}
\end{figure}

Figure \ref{fig:AudioFingerprintVerification} illustrates the verification process. The signed G-code is printed on a 3D printer of the same model used to create the fingerprint. The audio signal is recorded via a special mobile device application, which receives the signed G-code file, decrypts the MFTS and the eigenvectors and  compares the recorded audio signal to the MFTS as described in algorithm \ref{alg:audioFingerprintComparison}.

In order to verify the integrity of a new audio recording, we use a similar algorithm to extract the textual summary of the audio signal and compare it to MFTS. 

The algorithm receives three parameters: the signal of the new 3D printing recording , the MFTS, and the three eigenvectors associate with the MFTS.

%
%


\begin{algorithm}

	\begin{algorithmic}[1]
		\Function{AudioFingerprintComparison}{$signal, E, MFTS$}
		
		\State $TrimByMarkers(signal)$ \label{alg2:trim}
		\State $downsample = Resample(signal, 2000)$ \label{alg2:resampling}
		\State $S = spectrogram(downsample, 0.75, 0.1,1000, 20) $ \label{alg2:FFT}
		\State $S = S - mean(S)$ \label{alg2:mean}
		
		\State $afterPCA=S*E$ \label{alg2:project}
		\State $similarity  = cos(afterPCA, MFTS)$ \label{alg2:cos}
		\State $similarity  = smooth(similarity, 3)$ \label{alg2:smooth}
		
		\Return $similarity$
		
		\EndFunction
		
	\end{algorithmic}
	\caption{Audio Fingerprint Comparison}
	\label{alg:audioFingerprintComparison}
\end{algorithm}

\begin{description}
	\item[Line~\ref{alg2:trim} through \ref{alg2:mean}:] 
	The initial preparatory operations are identical to those in Algorithm \ref{alg:createMFTS}.
	
	\item[Line~\ref{alg2:project}:] 
	We next calculate the ${afterPCA}$ value as a projection of the spectogram on the eigenvectors that were determined in Algorithm \ref{alg:createMFTS}.	

	\item[Line~\ref{alg2:cos}:] 
	We then use cosine metrics to quantify the similarity the two vectors.
	Cosine similarity measures the cosine of the angle between vectors, meaning that identical vectors receive a score of 1; a 0 indicates that there is no correlation between the vectors. Vectors that are correlated in the opposite direction will be scored -1. For our case, lower similarity numbers indicate miscorrelation.
	At the end of this step we obtain a stream of similarity coefficients.

	\item[Line~\ref{alg2:smooth}:] 
	We apply a moving average filter to smooth out short-term acoustic fluctuations at the similarity stream output and to alleviate light pattern misalignments. Note that the smoothing filter span should match the desired resolution level in the verification processes. For instance, a short span is required for mismatch detection of fine printing movements but it would likely lead to an increase in false positives, especially in a noisy environment. In our experiments, we set the filter span to 3 (=0.3 sec.) 
	
\end{description}

Alignment is critical for this algorithm since the similarity is calculated by the cosine similarity of frames at the same time offset. Any misalignment might produce negative results. 
%
\\

\section{Experimental Evaluation}
\label{sec:ExperimentalEvaluation} 

In this section we present the experimental evaluation of the proposed algorithm. We discuss the setup, the modifications that were inserted into the 3D designs and the results (i.e the comparison graphs of the recordings).

\begin{figure*}
	\centering
	\includegraphics[width=.7\textwidth]{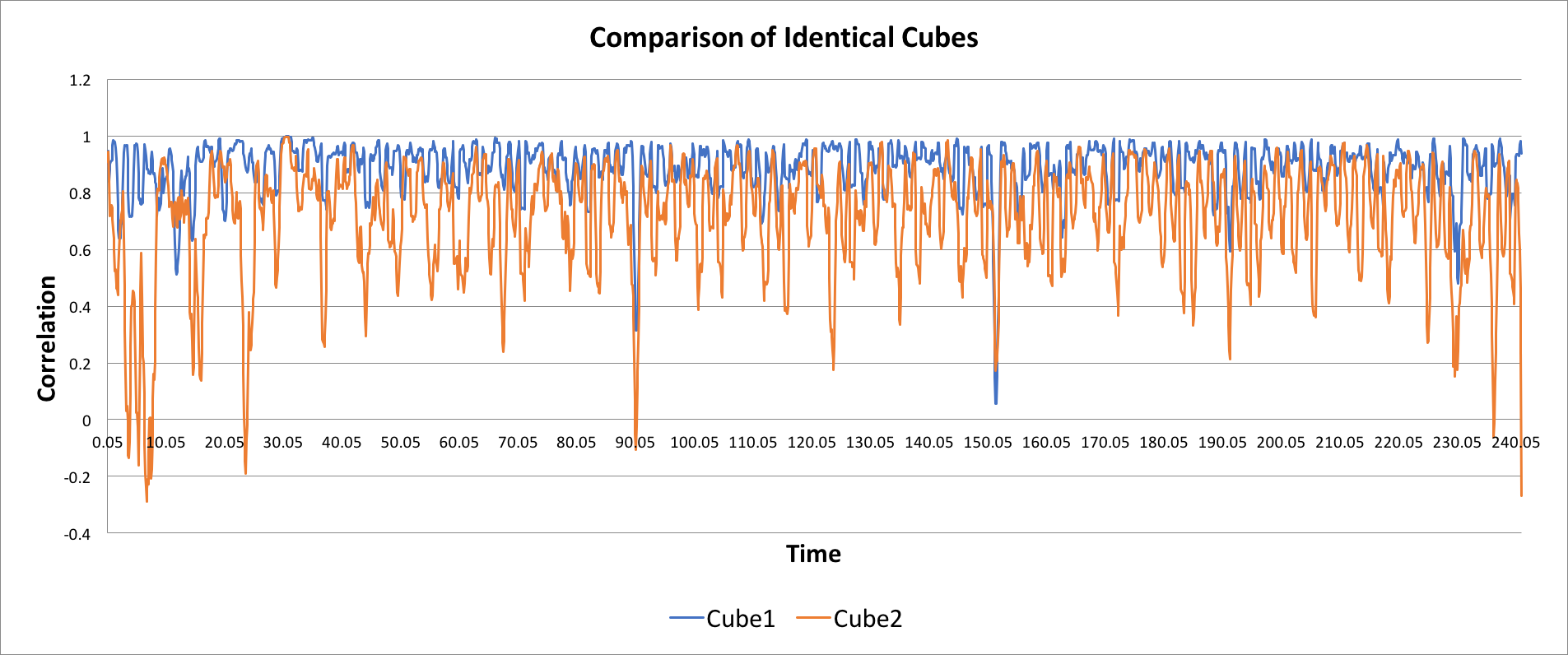}
	\caption{Audio comparison between 3D printing process of identical cubes (smoothing factor of 10)}
	\label{fig:identical_cubes}
\end{figure*}

\subsection{Experimental Setup}

All the experiments were performed in a lab environment. In the lab we used common PCs to create the G-code files that were copied to a SD card that was inserted into the 3D printer. 
We printed on a BCN3D Sigma printer that has independent dual extrusion (IDEX) technology, meaning that the 3D printer has 2 extruder heads that operate independently. The 3D printer runs the "BCN3D Sigma – Marlin" firmware and uses FDM technology. The maximum noise level this 3D printer reaches is 58 dB.

The software used for slicing the STL files is Cura-BCN3D, a version of the open source Cura software customized for the Sigma 3D printer. 
The design modifications were created by either changing properties in the Cura-BCD3D software or by modifying the G-code files. 

The audio recordings of the 3D printing process were taken using freeware applications on mobile devices. During the 3D printing process, the mobile device was placed adjacent to the 3D printer and recorded the audio in stereo at 44.1kHz.

\subsection{Experiments}

We began by generating the audio signature of a simple cube, then verified that the signature was consistent across 15 identical cubes. Then, for each modification type in figure~\ref{fig:result-table}, we have recorded 2 unmodified objects and at least 3 modified objects. The modified files were crafted in a decreasing significant of effects in order to determine the detection limits of this method. 
One of the audio recordings of the unmodified 3D object was used for the computation of master audio fingerprint (consist of $<E, MFTS>$ tuple generated by Algorithm \ref{alg:createMFTS}).
The other audio recordings were used in Algorithm \ref{alg:audioFingerprintComparison} to verify the detectability of modifications, and to determine false positive and negative rates.  

While the majority of the experiments were carried out on the cube geometry, we have validated them on rectangles, pyramids, and more complex geometries (e.g., a propeller).

\subsection{Experimental Results}

In this subsection we outline the experiments that were performed, focusing on determining normal and abnormal behavior and narrowing down on the algorithm's limitations. 
We've also recreated the modification that was presented in the dr0wned attack \cite{belikovetsky2017dr0wned} to test the proposed detection method.
We have recorded the audio signals of benign and modified propellers and compared their audio fingerprints. 
The algorithm has detected a significant deviation in the modified propeller. This is expected, since the modification was done on the STL file. Thus, the original and the modified STL files were sliced separately, creating different flows of G-code commands, resulting in dissimilar audio fingerprints.

Note that the graphs in this section are plotted with different smoothing factors. This is done for visual aid. The verification algorithm was calculated on data with the smoothing factor of 3.

\subsubsection{Normal and Abnormal Behavior}

In order to determine the tolerance of the algorithm, we've examined audio recordings of identical G-code files when compared to the pre-recorded audio master file. 
We have recorded 15 identical cubes in different settings and plotted the results of the comparison algorithm (algorithm ~\ref{alg:audioFingerprintComparison}).

Figure~\ref{fig:identical_cubes} depicts the output of algorithm \ref{alg:audioFingerprintComparison} (comparison of audio fingerprints) for two identical cubes. 
The plot for the first cube (Cube1) is representative for the vast majority of the tested audio recordings for unmodified 3D objects (12 out of 15); where the correlation value determined in Algorithm 2 concentrates around 0.8 - 0.9 values. 
The second cube (Cube2) represents a benign cube that causes a false positive detection. We can see that in the beginning of the recording there is a large plunge down in correlation. However, the graph syncs back. We have observed the re-syncronization always occurs on benign prints and is a indicator of integrity. However, this feature can mask some attacks that we will demonstrate later in this section. 

 From these results, we have learned about the normal behavior of the graph for comparing identical objects. We've seen that the correlation between the MFTS and each of the audio files in the test set is very high. For some of the compared files we can see periodic downward spikes that normally result from background noise, but they are rare and the correlation returns to higher values. 

\begin{figure*}
	\centering
	\includegraphics[width=.7\textwidth]{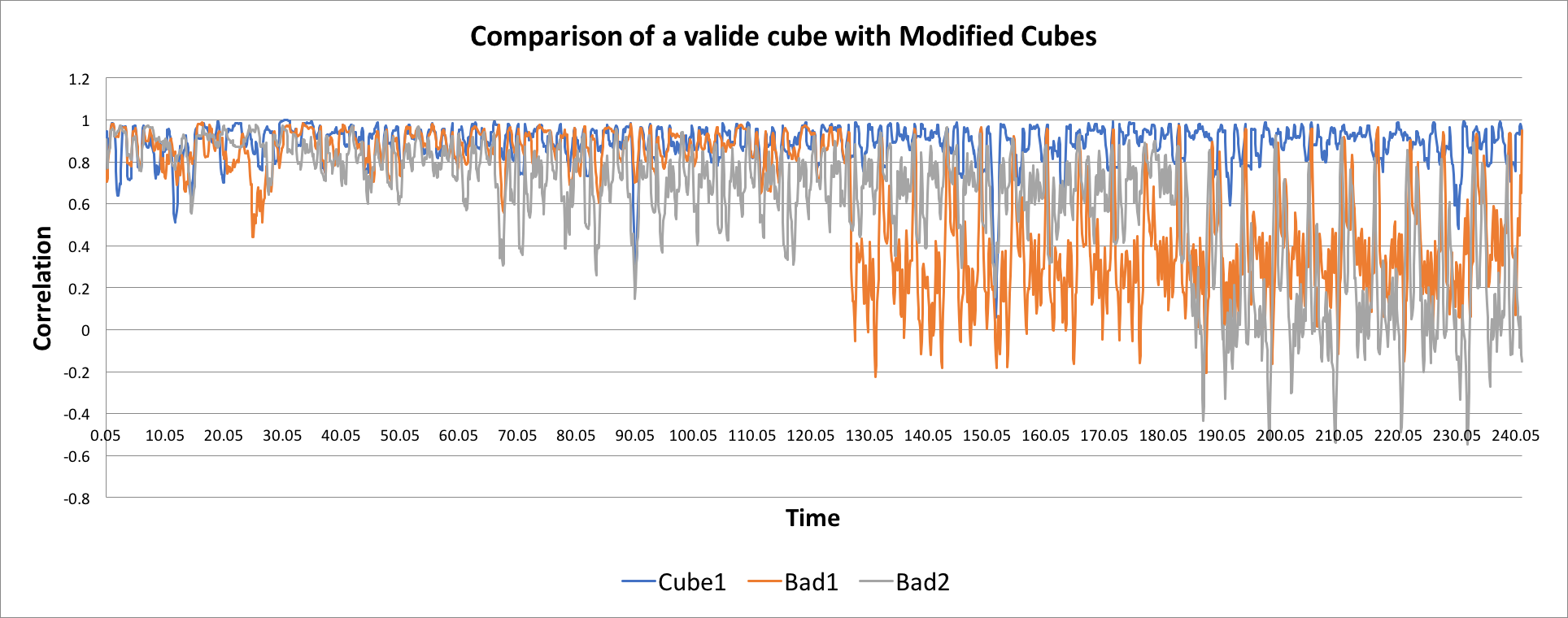}
	\caption{Audio comparison between valid and modified 3D prints (smoothing factor of 10)}
	\label{fig:comparison_valid_mal_cubes}
\end{figure*}

When comparing against modified objects, it is possible to detect not only a deviation from the expected pattern but also the exact point when the first modification occurs. In figure~\ref{fig:comparison_valid_mal_cubes} we can see the fingerprint comparison between the benign ``Cube1'' recording and recordings of two modified cubes. 
The similarity of Cube1's recording to the MFTS is high throughout the whole manufacturing process; the only deviations are sparse spikes of reduced correlation. 
The audio recordings of the 3D printing process of the modified cubes lose synchronization exactly when the modification of the G-code instruction sequence occurs. The ``Bad1'' cube's G-code contains two dummy moves at layer
20 out of 40. We can see that the graph loses synchronization at about the middle of the recording. In the cube marked ``Bad2'', we've removed a single G1 printing command at layer 30. We can see that the graph loses synchronization towards the last quarter of the recording.

\subsubsection{Detection Limits}
\label{subsec:detection_limit}



\begin{figure}[H]
	\centering
	\includegraphics[width=.4\textwidth]{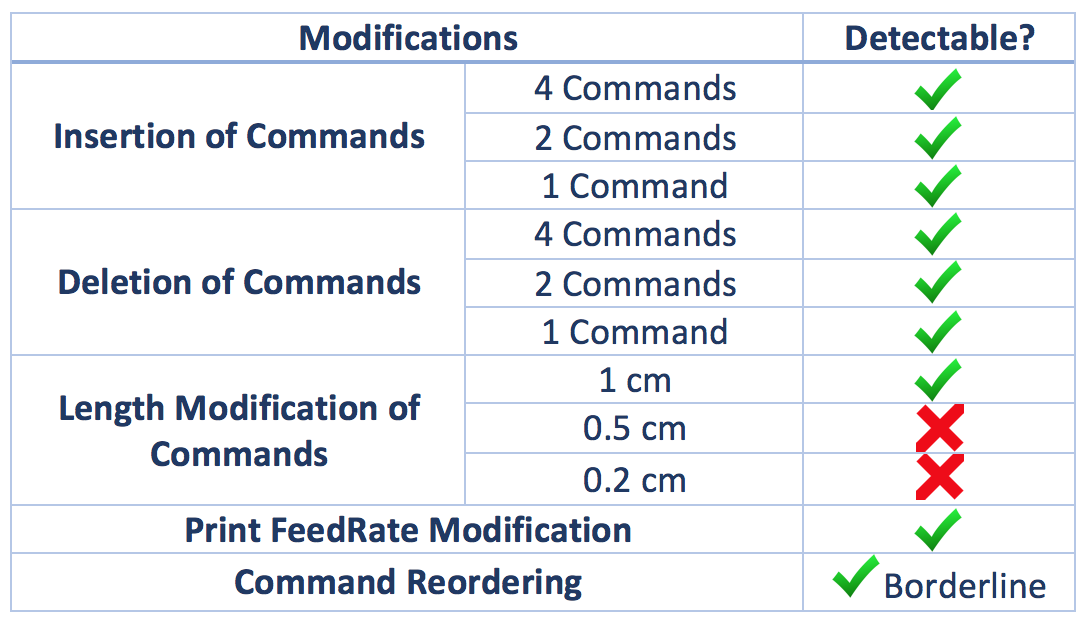}
	\caption{Summary of the detection results on various modifications}
	\label{fig:result-table}
\end{figure}

Determining the detectability limit of this technique relies on the minimal malicious deviations that can be introduced. 
Attacks discussed in the literature are modifications of the 3D object geometry~\cite{sturm2014cyber, zeltmann2016manufacturing, chhetri2016kcad, belikovetsky2017dr0wned}, 3D object orientation during the printing process~\cite{yampolskiy2015security, zeltmann2016manufacturing}, and manipulations of the manufacturing process~\cite{yampolskiy2015security}. All these attacks result in changes to tool path instructions.
Attacks involving malicious firmware can deliberately misinterpret G-Code commands~\cite{moore2017implications}, but these manipulations result in actual tool path changes; these can be described (and tested) as changes introduced into the G-code.
Therefore, for the evaluation of the proposal's limitations, we've tested changes at the individual G-Code command level. Table \ref{fig:result-table} summarizes the results of those experiments. 
More specifically, we've tested the following modifications: 

\begin{enumerate}
	\item Insertion of additional G-Code commands
	\item Deletion of G-Code commands included in the tool path of a benign 3D object
	\item Modification of parameters to an individual movement command along one axis 
	\item Modification of the extruder's speed  
	\item Reordering of G-code commands
\end{enumerate}

To analyze the detection thresholds, we've performed controlled tests on sections of 20 seconds of audio recordings. We used a cube due to its repetitive geometry; each section translates to exactly 4 layers in the cube. Each section was marked with the audio marker 
and the modification was inserted in the $3^{\text{rd}}$ layer, {\em i.e.,} in the second half of the recording. 

For every modification type, we've tested decreasing levels of deviations and 
validated whether or not such deviations can be detected with the proposed approach.

The comparison was performed with a smoothing factor of 3; a lower factor would improve the detection resolution while increasing the false positive rate.

\paragraph{Insertion of Commands}

\begin{figure}[H]
	\includegraphics[width=.3\textwidth,height=.13\textheight]{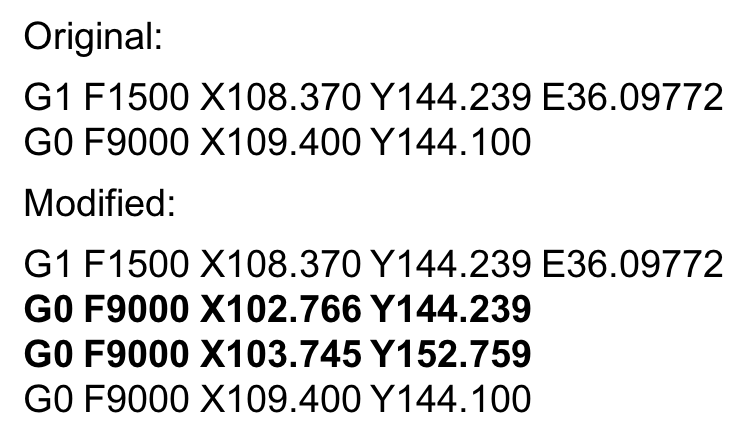}
	\caption{Insertion of two G0 moves in G-code}
	\label{fig:gcode-insert}
\end{figure}

\begin{figure}[!htb]
	\centering
	\includegraphics[width=.45\textwidth]{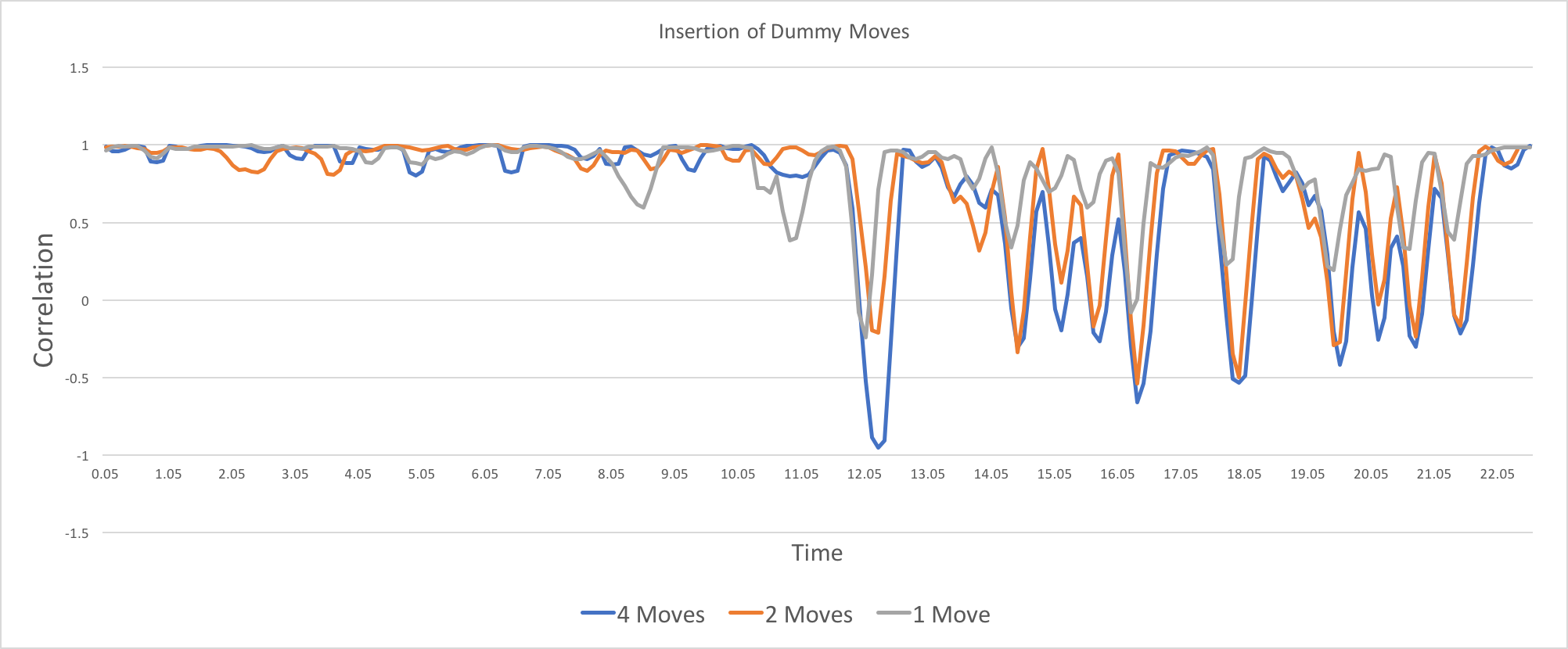}
	\caption{Audio comparison of original cube versus three modified cubes. The modification: insertion of G0 moves}
	\label{fig:gcode-insert-graph}
\end{figure}

We've inserted additional G0 commands 
into the manipulated G-code files. 
The G0 command translates to a movement of the extruder to the specified X and Y coordinates without extruding the filament. 
Figure~\ref{fig:gcode-insert} shows the original and the modified G-code commands of one file; two additional G0 commands were inserted. 
Figure~\ref{fig:gcode-insert-graph} shows the similarity graph comparing the 3D printing process of three modified G-code files and the audio master file. 
They contained four, two and a single inserted G0 command, respectively.
The addition of G0 commands desynchronizes the prints, and consequently the degree of similarity 
degrades dramatically right after the execution of the inserted commands.

\paragraph{Deletion of Commands}

\begin{figure}[!htb]
	\includegraphics[width=.3\textwidth,height=.13\textheight]{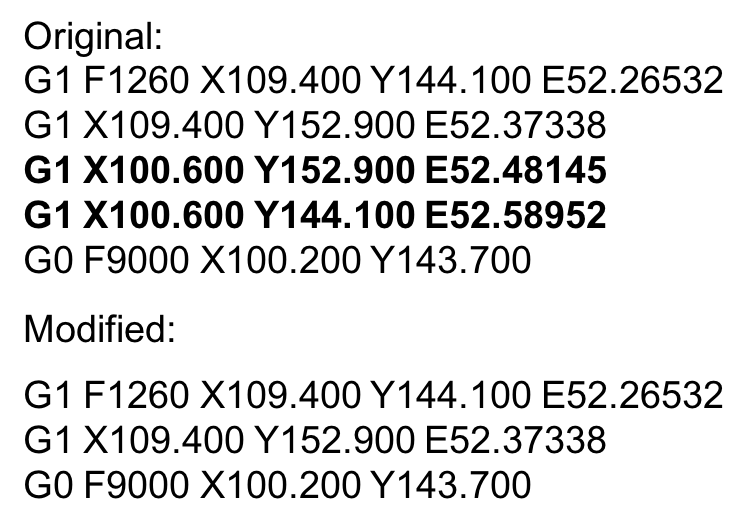}
	\caption{Deletion of two G1 print moves in G-code }
	\label{fig:gcode-delete}
\end{figure}

\begin{figure}[!htb]
	\centering
	\includegraphics[width=.45\textwidth]{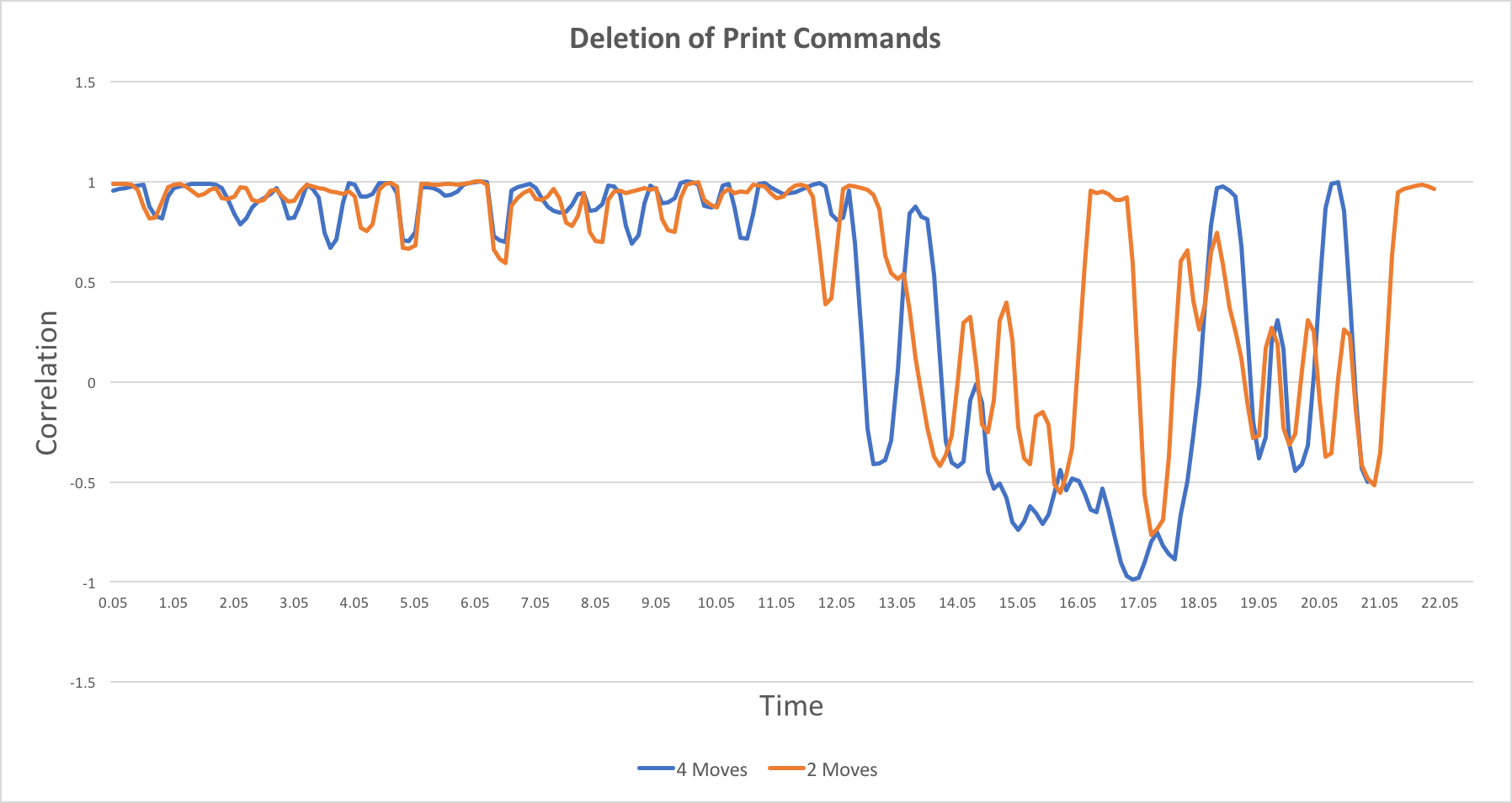}
	\caption{Audio comparison of original cube versus two modified cubes. The modification: deletion of G1 print moves}
	\label{fig:gcode-delete-graph}
\end{figure}

The G1 command moves the extruder to a specified (X, Y) coordinate while extruding filament.
Changes to the 3D object geometry (both internal and external) will likely involve modifying G1 commands.
To validate the detectability of such changes, we've deleted G1 commands from the G-code files.
Figure~\ref{fig:gcode-delete} shows both the original and the modified G-code commands on one file. 
Figure~\ref{fig:gcode-delete-graph} shows the similarity graph comparing the 3D printing process of the unmodified and two modified files. 
They had four and two removed G1 commands, respectively. Even the deletion of a single G1 command breaks synchronization; this is reflected by the dramatic degradation of the degree of similarity right after the removed G-code command in line "Bad2" in figure \ref{fig:comparison_valid_mal_cubes}.

\paragraph{Modification of Movement Length on the Axis}

We have tested several modifications to movements along the axes, including extended and shortened move length. 
For both we successively reduced the length of the deviation from the original command and observed the impact on the similarity plot. 

The minimum change which still broke synchronization at our smoothing factor was a modification of 1 cm in length on a single G1 print command. 
This break is similar to inserting or deleting a G1 print command in figure ~\ref{fig:gcode-insert-graph}. 
Modifications of 0.5 cm and 0.2 cm also break the synchronization, but the degree of similarity does not degraded as drastically.

\begin{figure}[!htb]
	\includegraphics[width=.3\textwidth,height=.13\textheight]{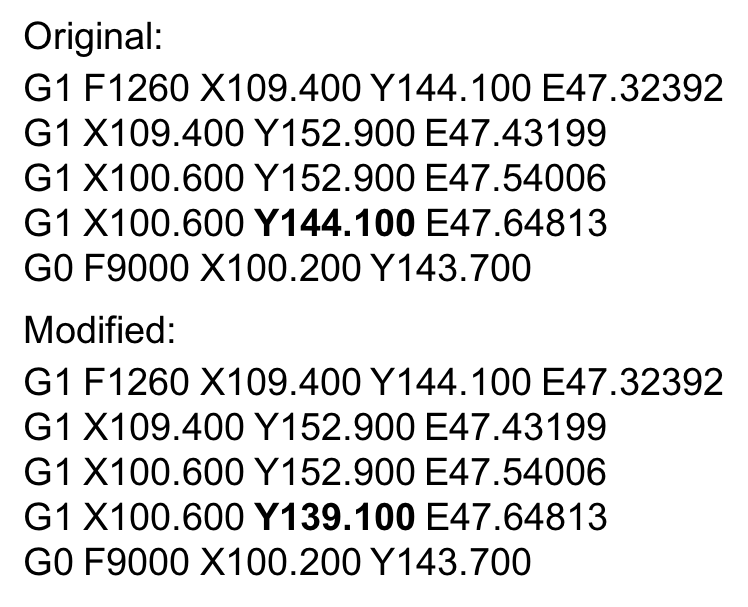}
	\caption{Extend a single G1 print move in G-code}
	\label{fig:gcode-ymore}
\end{figure}

\begin{figure}[!htb]
	\centering
	\includegraphics[width=.45\textwidth]{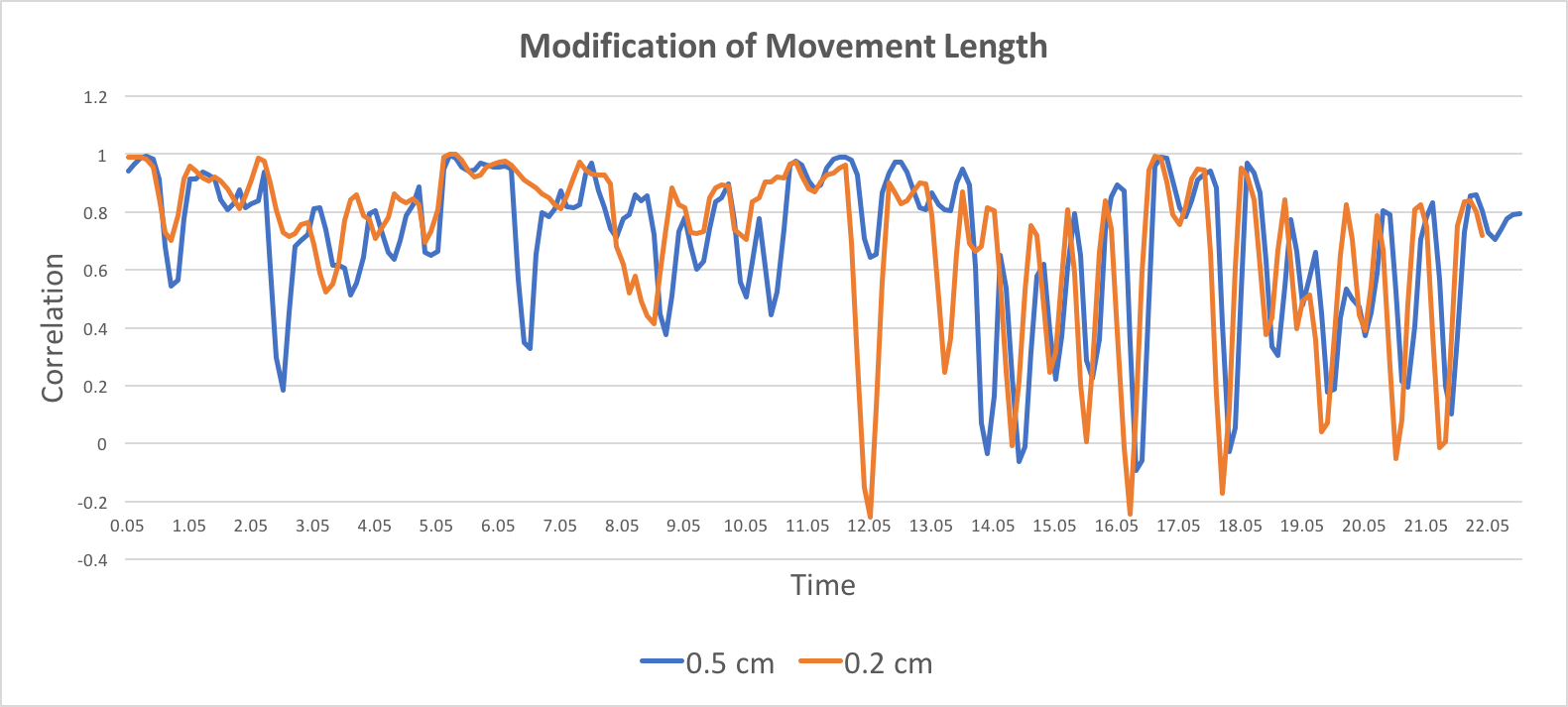}
	\caption{Audio comparison of original cube versus a modified cube. The modification: extending the movements on one of the axis.}
	\label{fig:gcode-ymore-graph}
\end{figure}

Figure~\ref{fig:gcode-ymore}
shows the original and the modified G-code commands; the modification extends the Y-axis movement by 0.5 cm.
Figure~\ref{fig:gcode-ymore-graph} shows the similarity graph  between the original audio file and the recordings of two modified G-code files with the extensions of 0.5 cm and 0.2 cm respectively.
The experiments have been performed for all three axes, and present uniform detectability thresholds.

\paragraph{Modification of Extruder's Speed}

The amount of the filament deposited during a movement is a function of the speed of the nozzle movement and the speed of the filament extrusion motor (both controlled by G-code command G1) .
Modifying the feedrate parameter in G1 commands changes the speed of the executed move. 

In this case it is more difficult to state the minimal change needed to break synchronization since there are two factors involved: the length of the move and the original feedrate of the command. 
We saw that we can break syncronization by slowing down the travel speed of two G1 print commands (length move of 1 cm).

\begin{figure}[!htb]
	\includegraphics[width=.3\textwidth,height=.13\textheight]{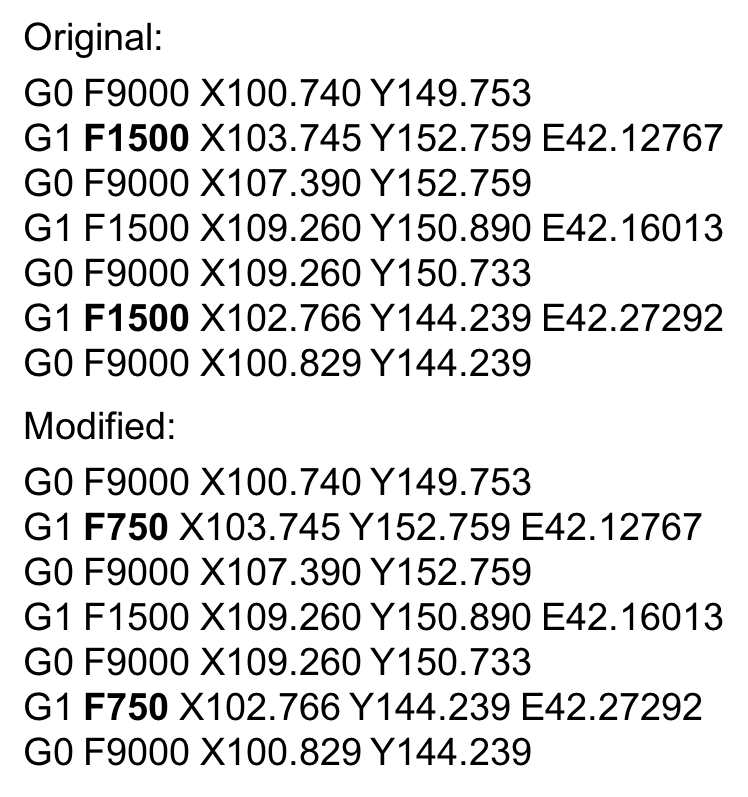}
	\caption{Extend a single G1 print move in G-code}
	\label{fig:gcode-f}
\end{figure}

\begin{figure}[!htb]
	\centering
	\includegraphics[width=.45\textwidth]{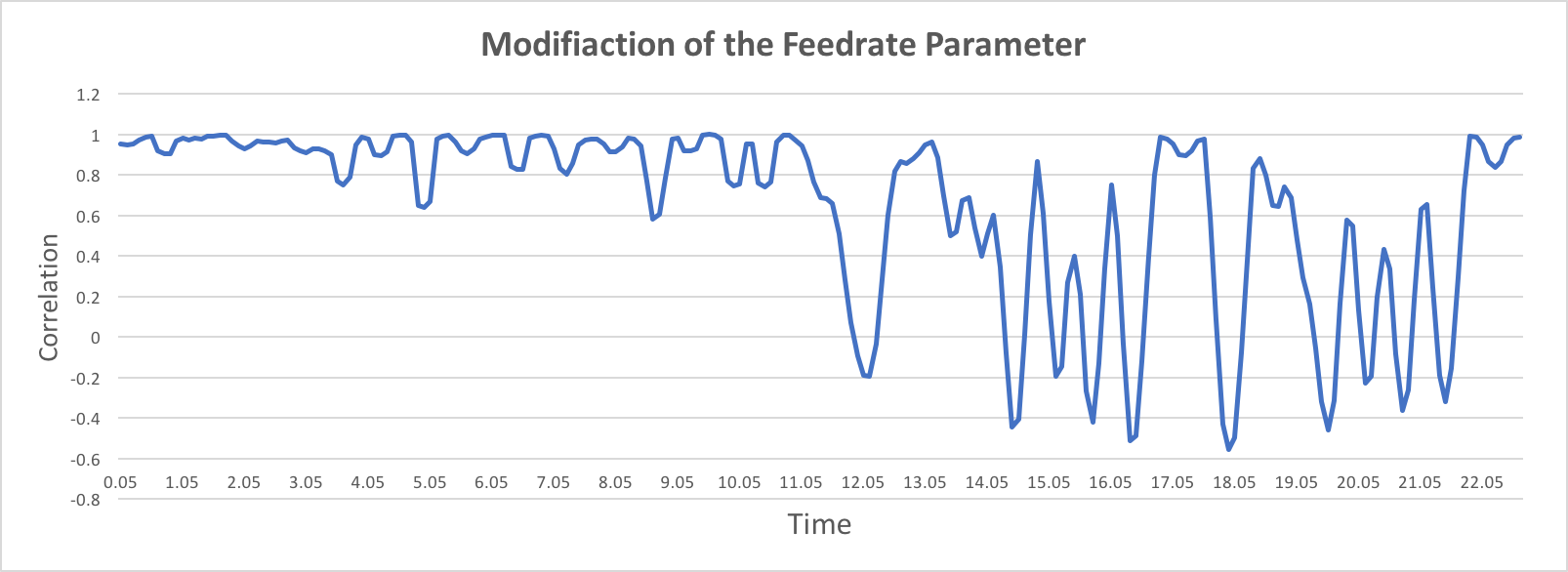}
	\caption{Audio comparison of original cube versus a modified cubes. The modification: changing the feedrate parameter of move commands }
	\label{fig:gcode-f-graph}
\end{figure}

We have performed experiments to identify the threshold at which changes of the extruded filament can be detected.
Figure~\ref{fig:gcode-f} shows the original and the modified extrusion speed parameter of G-code command G1. 
Figure~\ref{fig:gcode-f-graph} shows the similarity graph generated for the modified G-code file containing a feedrate change on two G1 commands by half (1500 to 750).

\paragraph{Reordering of G-code commands}

\begin{figure}[!htb]
	\includegraphics[width=.3\textwidth,height=.13\textheight]{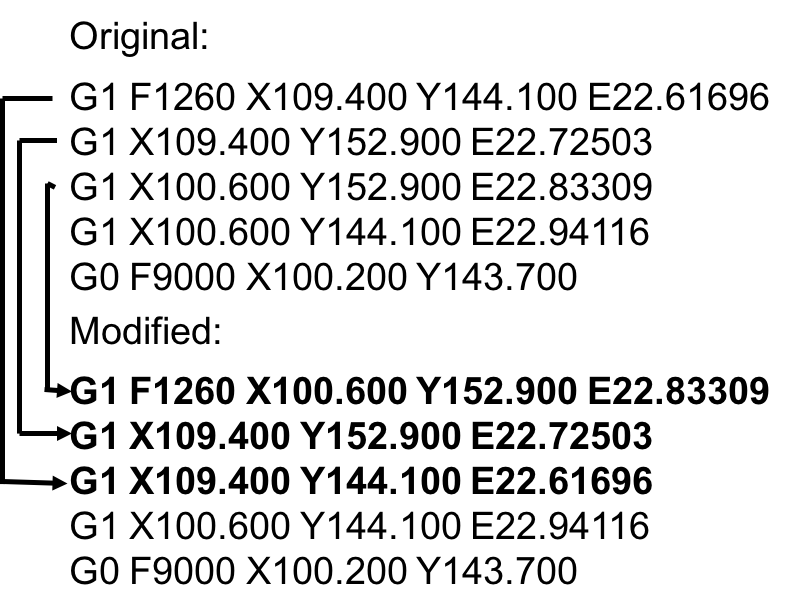}
	\caption{Reorder 3 G1 print commands in G-code}
	\label{fig:gcode-reorder}
\end{figure}

\begin{figure}[!htb]
	\centering
	\includegraphics[width=.45\textwidth]{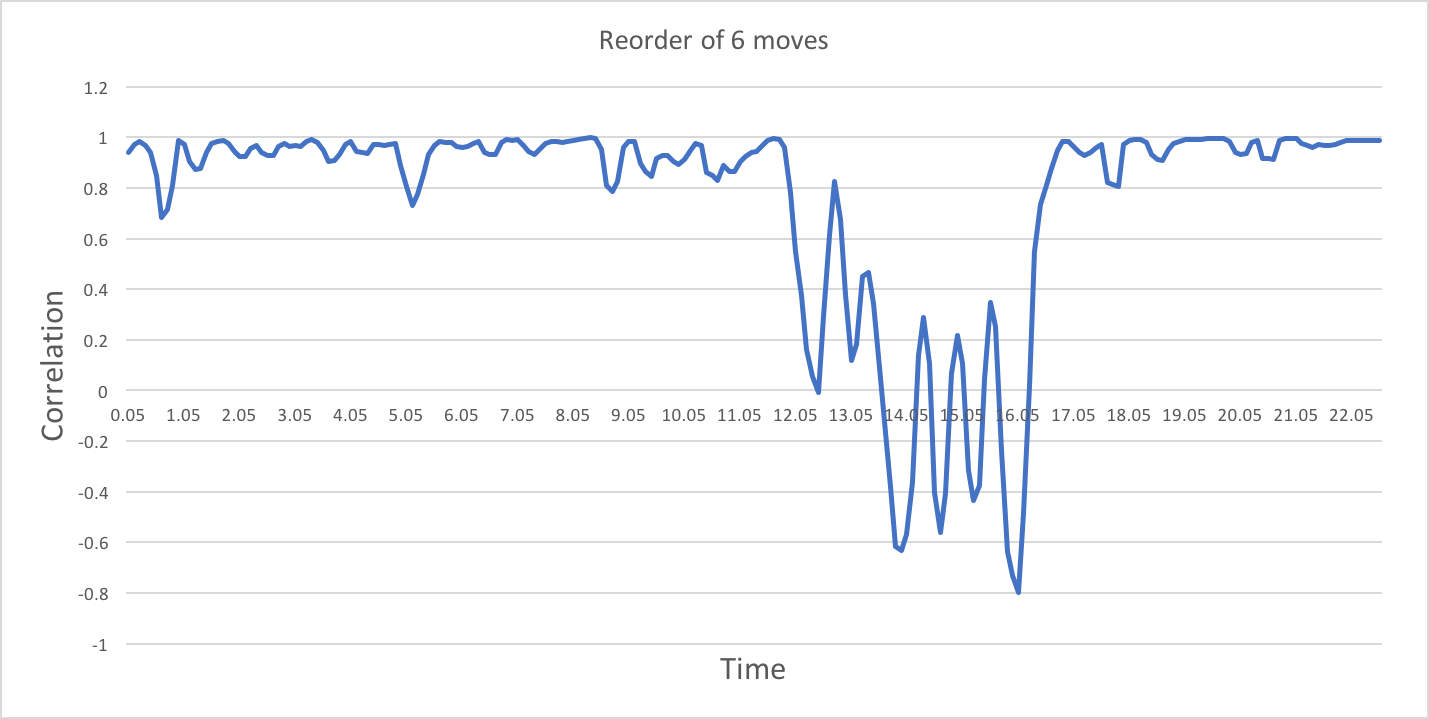}
	\caption{Audio comparison of original cube versus a modified cube. The modification: reorder of 3 G1 print commands }
	\label{fig:gcode-reorder-graph}
\end{figure}

\begin{figure}[!htb]
	\centering
	\includegraphics[width=.45\textwidth]{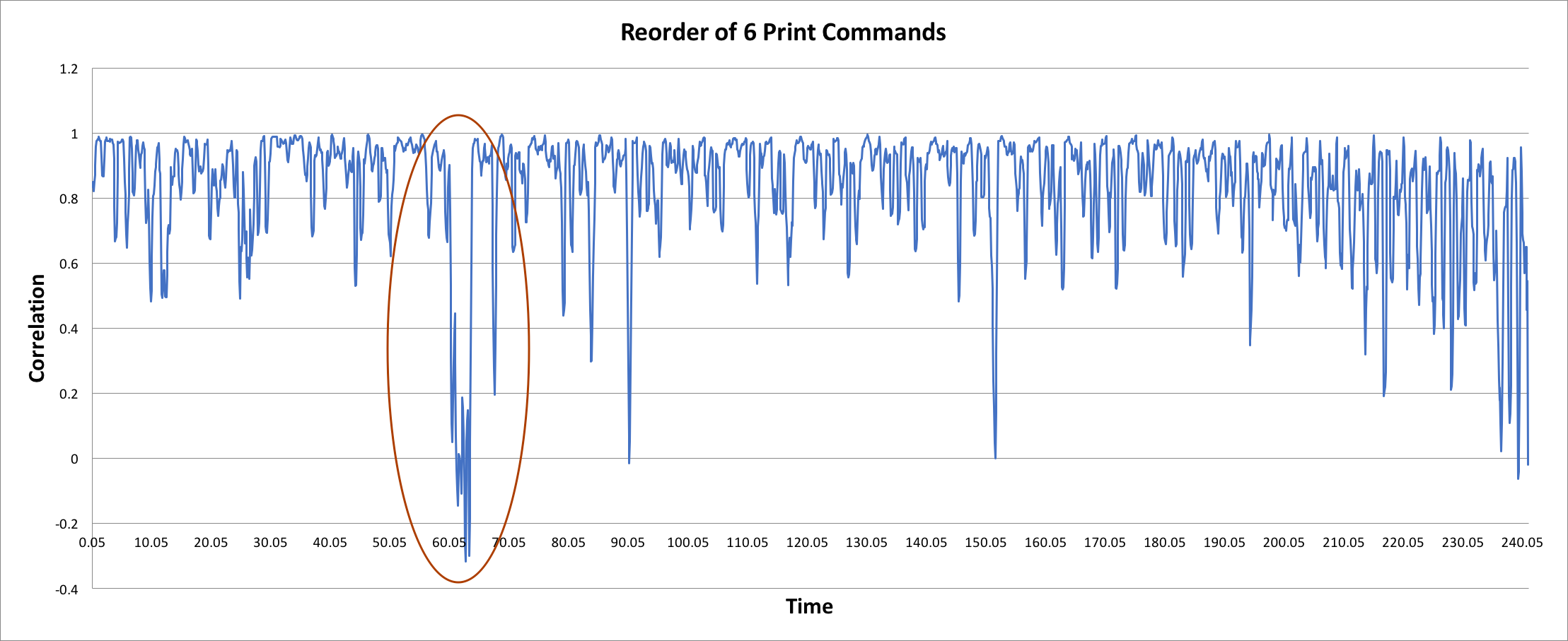}
	\caption{Audio comparison of original cube versus a modified cube. The modification: reorder of 6 G1 print commands in layer 8 out of 50 (smoothing factor of  5)}
	\label{fig:bigfile-reorder-graph}
\end{figure}

Reordering G-code commands does not modify the geometry of the object, but might effect the quality of the object. 
However, reordering a few commands does not appear to hurt the overall syncronization. 
Figure~\ref{fig:gcode-reorder} contains the modified G-code. 
For one of the 3D printed cube's layers, the new G-code file contains the same commands but in a different order. 
The similarity plot in Figure ~\ref{fig:gcode-reorder-graph} 
shows that this causes a disturbance while the reordered commands are executed. 
However, since the overall timing has not changed, the line syncs back up afterwards (layer 4 out of 4). When we examine the effect of the reordering on the comparison graph of the entire cube (see Figure~\ref{fig:bigfile-reorder-graph}), we can see a noticeable disturbance near the beginning of the file but none afterward. We conclude that the reordering of commands will be noticeable but difficult to distinguish from disturbances introduced by background noise. 

\subsubsection{Testing Disturbances} 

There are several factors that can influence the signal of an audio recording. 
In order to test the resilience of the proposed solution to factors like recording device, microphone position, and background noise, we performed several audio recordings while introducing disturbances. 
Below we describe the tested disturbances.

\begin{figure}[!htb]
	\centering
	\includegraphics[width=.45\textwidth]{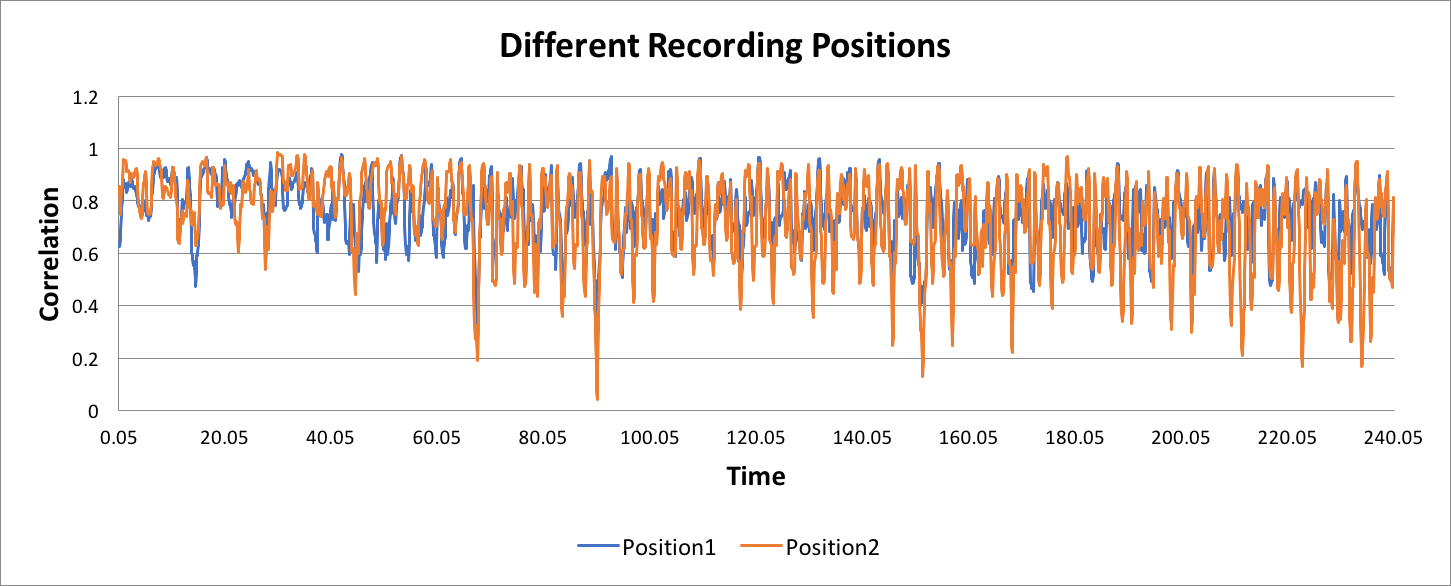}
	\caption{Audio comparison graph of cubes recorded from different positions and by different mobile device than the audio master file (smoothing factor of 10)}
	\label{fig:recordingpositions}
\end{figure}

\paragraph{Different Recording Devices}
The audio signal would be recorded by different applications on different mobile devices. In Figure ~\ref{fig:recordingpositions}, the recordings were taken by a different mobile device than the master file.

\paragraph{Recording Positions} 
Part of the experiments tested different recording positions near the 3D printer. While the audio master file was recorded with the microphone at the left side of the 3D printer above the extruder, two further audio recording were taken with the microphones at the right side of the printer and in the front, about 20 cm below the extruder head. 
Figure~\ref{fig:recordingpositions} show similarity plots for audio recordings taken at different locations. 

\paragraph{Background Noise}
The recordings were done in a lab environment, some at night in a quiet environment, others during daytime with mild background noise. The effects introduced by the noisy environment appear as short plunges in the similarity graph. 
These are limited to the duration of the noise; the background noise does not affect synchronization of audio recordings. 
In an extremely loud environment with permanent background noise, this behavior might cause false positives.

\begin{figure}[!htb]
	\centering
	\includegraphics[width=.45\textwidth]{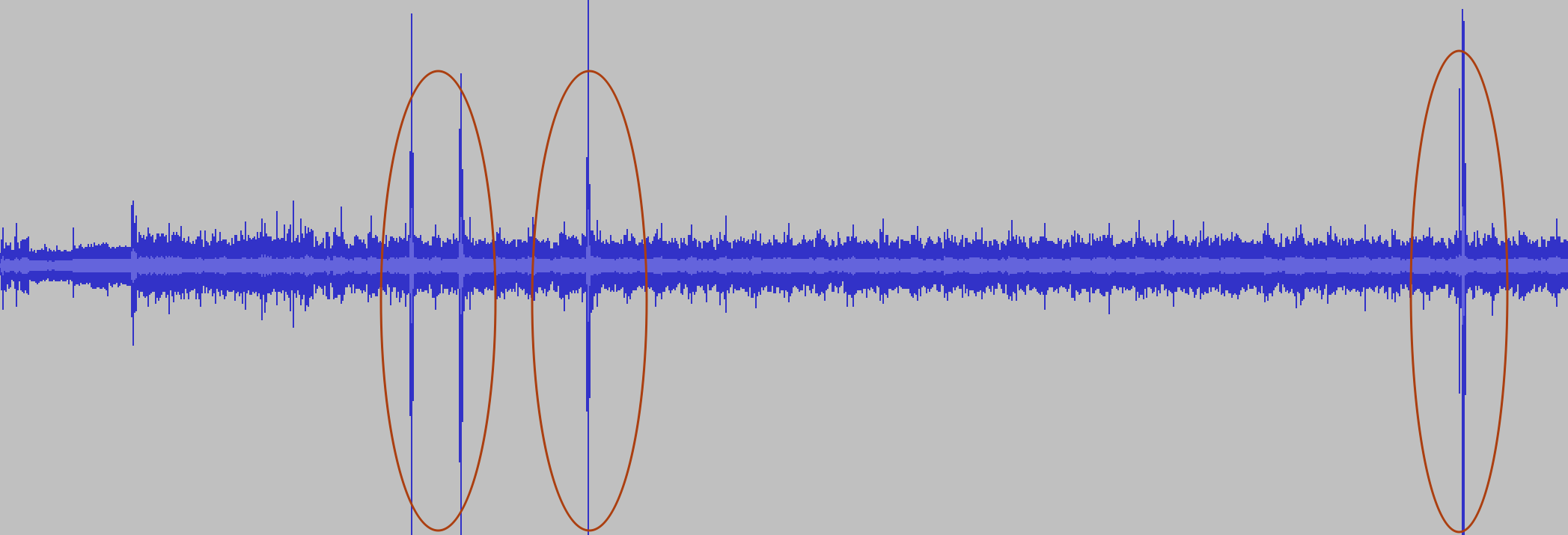}
	\caption{Audio signal of a recording of a cube with introduced loud background noises}
	\label{fig:noise-signal}
\end{figure}

\begin{figure}[!htb]
	\centering
	\includegraphics[width=.45\textwidth]{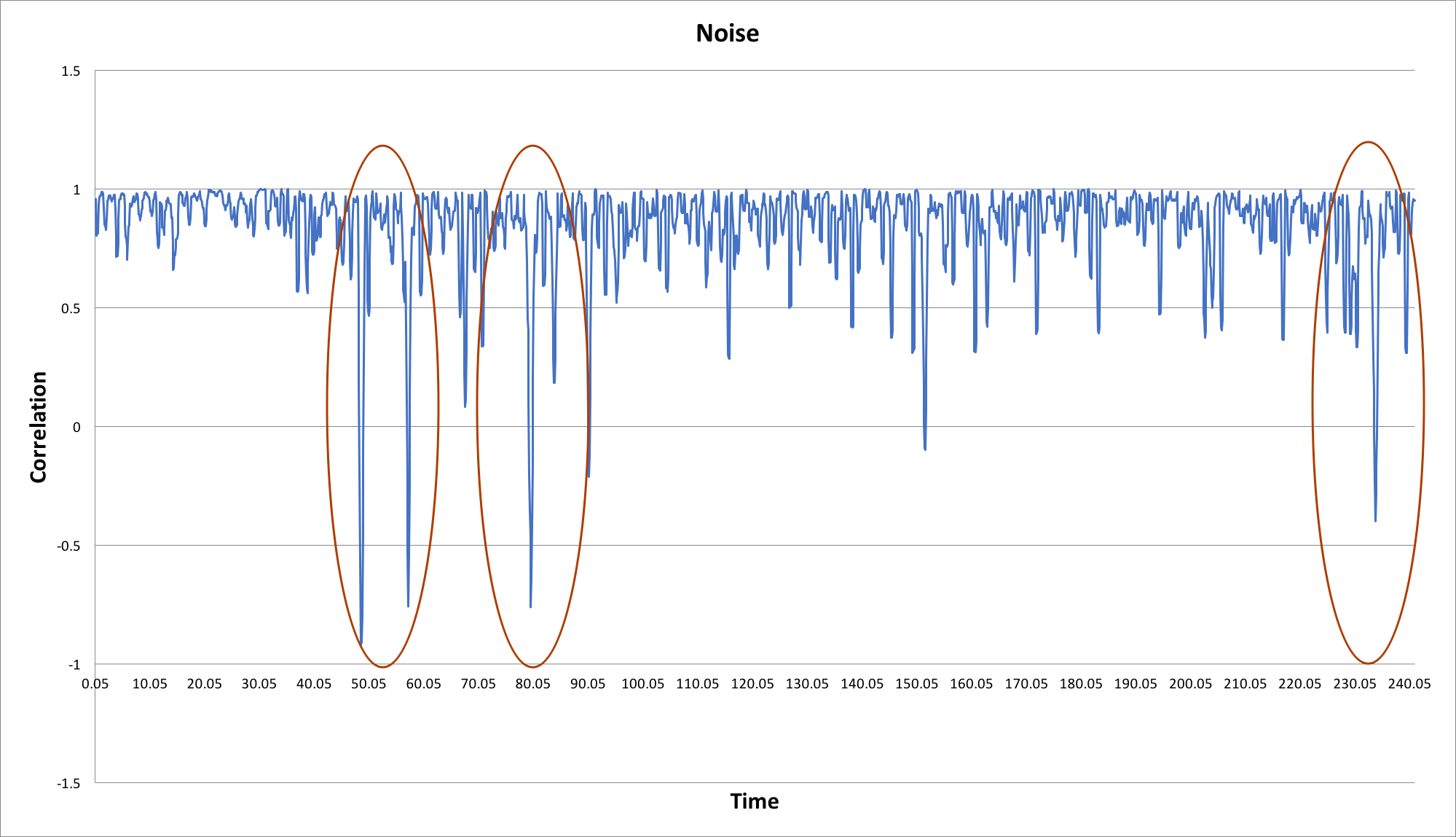}
	\caption{Audio comparison of original cube and an identical cube that was recorded with introduced background noises (smoothing factor of 5)}
	\label{fig:noise-graph}
\end{figure}

%

\subsection{Algorithm Limitations} 

\begin{figure}[!htb]
	\includegraphics[width=.3\textwidth,height=.13\textheight]{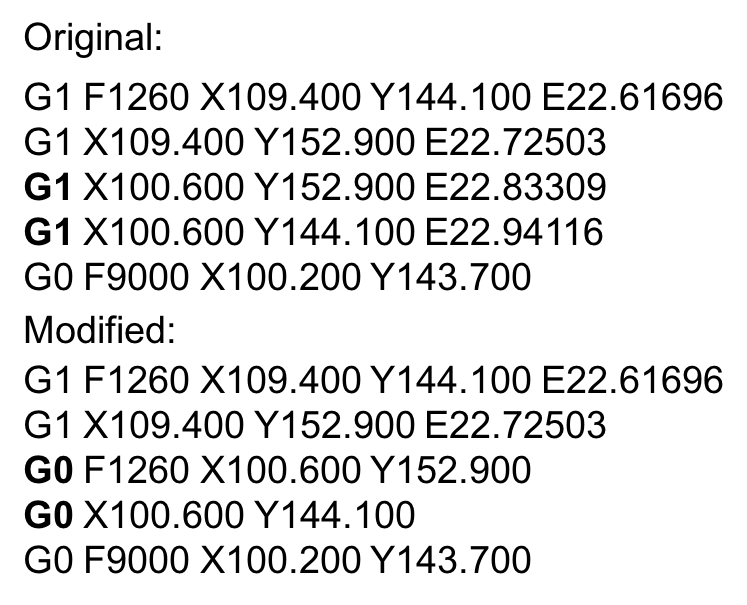}
	\caption{Replace 2 G1 commands with G0 commands in G-code}
	\label{fig:gcode-replacemove}
\end{figure}

\begin{figure}[!htb]
	\centering
	\includegraphics[width=.45\textwidth,height=.10\textheight]{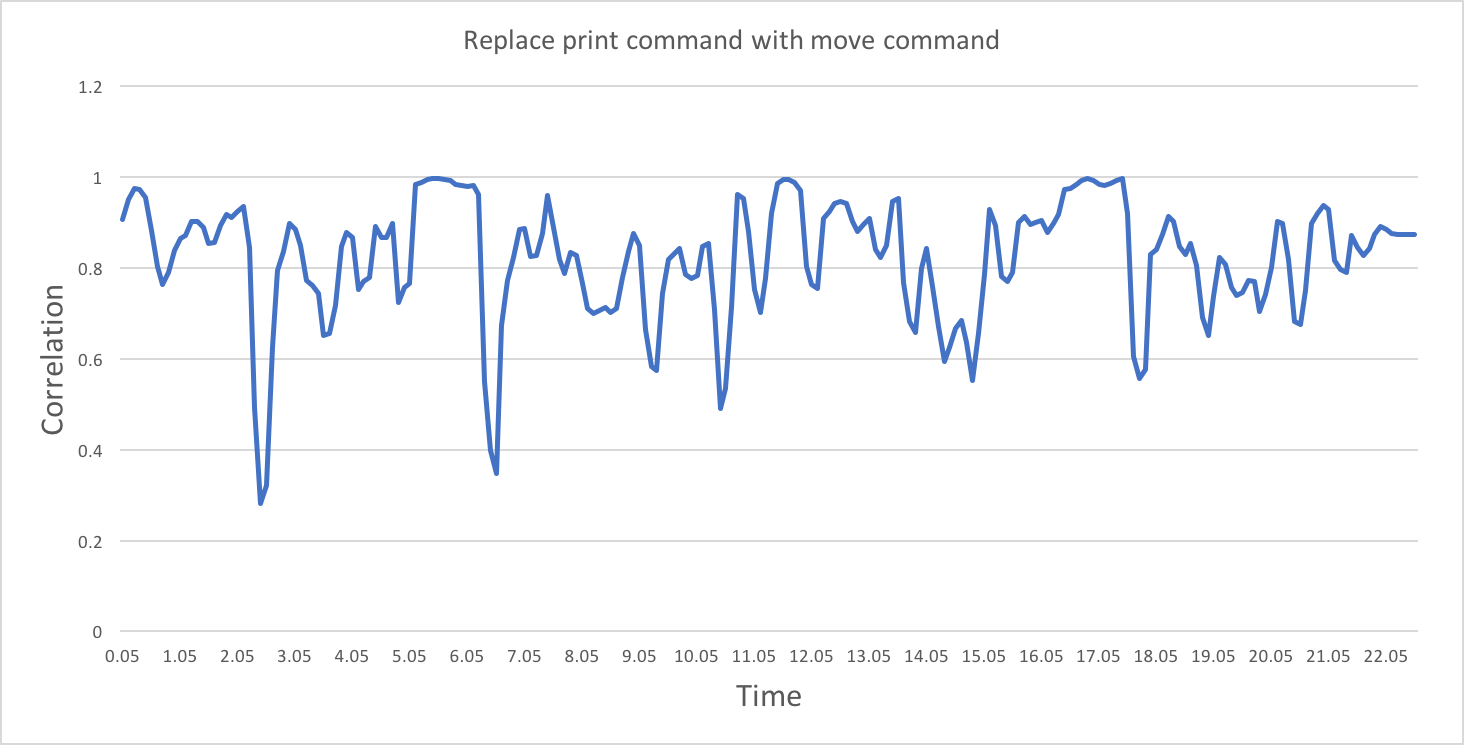}
	\caption{Audio comparison of original cube versus a modified cube. The modification: Replace 2 G1 commands with G0 commands }
	\label{fig:gcode-replacemove-graph}
\end{figure}

Since pattern similarity is calculated on a frame-by-frame basis, the algorithm relies on time synchronization. Modifications that cause a momentary mismatch but do not break the overall syncronization might therefore be interpreted as false positives. 
The main limitation we have discovered is in detecting command replacements of identical length, {\em e.g.,}
 replacing G1 print commands with dummy G0 move commands with the same frequency.
The move still takes place, but without filament extrusion. Since this does not affect the subsequent synchronization, the audio difference is momentary. We have tested the replacement of two G1 print moves with G0 moves of the same feedrate (figure ~\ref{fig:gcode-replacemove}). The audio recording of this modification is highly correlated with the audio master file as shown in figure ~\ref{fig:gcode-replacemove-graph}. 

\subsection {Modification Indicator}

\begin{figure}[!htb]
	\centering
	\includegraphics[width=.45\textwidth]{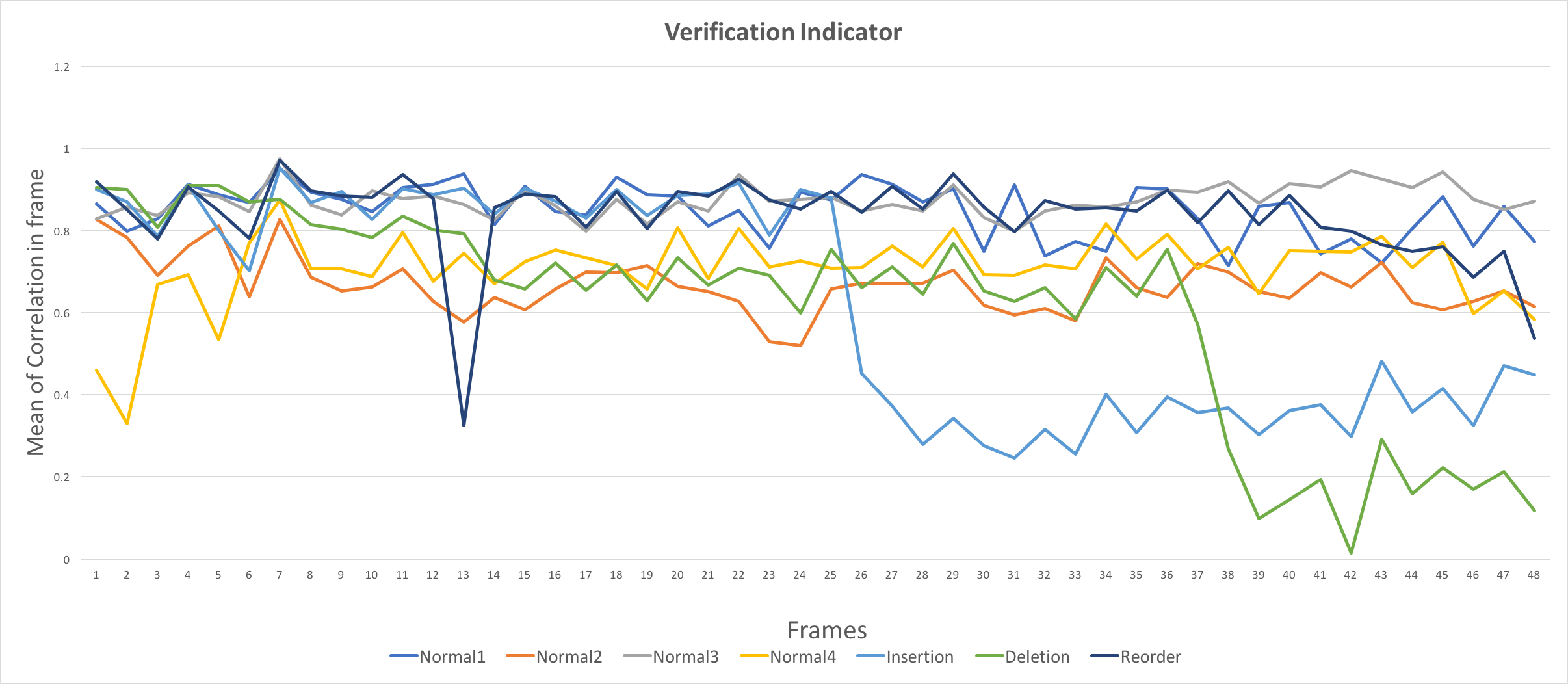}
	\caption{Plot of the mean value for time frames of 10 seconds on selected prints }
	\label{fig:verification-indicator}
\end{figure}

To detect modifications, we have looked for significant changes in the mean value of the signal. To eliminate large, brief dips in the correlation value, many of which are due to background noise, we apply a large smoothing factor to the similarity graph.  
We have calculated the mean value for time frames of 5 seconds and looked for sections where the graph decreases overall at least 0.4 points for 4 consecutive time frames. 
Figure \ref{fig:verification-indicator} shows false positive and negative examples. Most of the identical cubes do not pass this value, except for cube "Normal4" that starts out of sync. 
For the modified cubes, we can see that there are rapid significant decreases in the graph, which pass the threshold and are flagged as malicious. The cube marked "Reorder", which contains 6 reordered print commands, gives a borderline result, and is barely detected by this indicator. 

%
%
%
%


\section{Signature Verification Use Cases}
\label{sec:use_cases}

\begin{figure}
	\centering
	\includegraphics[width=.25\textwidth,height=.1\textheight]{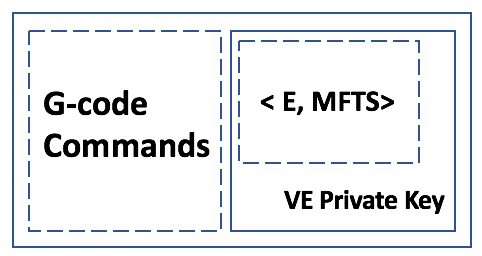}
	\caption{Signed G-code file format}
	\label{fig:signedgcodeformat}
\end{figure}

In this section we discuss how the audio fingerprint is attached to the signed G-code file and the verification process. Then, we present several use cases for this verification method. 

After running the fingerprint generation algorithm (algorithm \ref{alg:createMFTS}), we have a tuple of the eigenvectors that were calculated for the original audio signal and the MFTS. This is encrypted using the private key of some Verification Entity (VE), who varies depending on scenario. This data is appended as a comment to the G-code file as illustrated in figure \ref{fig:signedgcodeformat}.

Since the verification algorithm is sequential, verification can be done in real-time. 
To facilitate this, the VE has to release a mobile device application that contains its public key. The application receives the signed G-code file, decrypts the second part of the file using the VE's public key and starts recording the audio signal of the new printing process. 
It monitors for the start marker and then for every frame in the audio, it calculates the signature and compares it to the corresponding frame in the MFTS.

This real-time detection method indicates a possible deviation from the audio master file exactly at the point of the modification, thus saving time and material. 

We see several application scenarios of the proposed manufacturing process verification. 
In the consumer market, a custom app for a mobile phone could implement both algorithms. An desktop 3D printer user could use this app to record a 3D printing process and mark it as a ``master fingerprint.'' When the same 3D object is printed on the same equipment, the app can record and verify the manufacturing process. 

A drawback of this approach is that a consumer end-user is unlikely to be able to verify whether the first printed object was compliant to the original blueprint file. However, the end-user can still rely on their own experience of the first object, {\em i.e.,} whether or not the object's quality was sufficient for the intended application.
In industrial settings the same application could be used and, after recording of a master audio fingerprint, the manufactured object can be verified with non-destructive and/or destructive methods.

Another use case is a possible tamper-proof verification that an OEM can offer as a service. 
Here we use as an example the high-end Voxel8 FDM 3D printer. Instead of using a local PC to slice and to translate the 3D object's blueprint into a tool path, an end-user is submits the blueprint to a cloud service; then, when the user prints the design, translation operations are performed by the cloud service and a tool path is sent to the 3D printer via network.
It is easy to imagine that such an OEM can integrate an audio recording device into the 3D printer. After a 3D object is printed first time, a master fingerprint can be generated and stored in the cloud. Every time an end-user prints the same model, the audio recording can be verified against the stored master fingerprint; in case of deviations, the end-user can be notified that the printed object might have been tampered with. We see this as an attractive service that might be offered by OEMs.
In industrial settings, such as automotive shops, this approach would ensure the necessary quality of the part; furthermore, a cloud-based approach might be attractive to prevent IP infringement.

We also consider to peer-verification of printed objects using the proposed approach. 
In examples like an auto shop, multiple users will print the same parts at different locations. We assume that master files can be recorded for all printers, in order to accommodate possible deviations in mechanical components and resulting discrepancies in sound. Nevertheless, master audio fingerprints from different sites could be peer-verified against each other, within certain tolerances. This approach could identify compromised sites, with master audio fingerprints dissimilar to their peers.

\section{Related Work} 
\label{sec:related_work}

THe security of Additive Manufacturing is a research area pioneered around 2014~\cite{xiao2013security, sturm2014cyber, yampolskiy2014towards, yampolskiy2014intellectual, depoorter2013intellectual, holbrook2014digital}. 
So far, two major threat categories 
have been identified for AM: (1) sabotage and (2) violation of Intellectual Property (IP). 
Sabotage attacks aim to inflict physical damage, e.g., by compromising part quality or by damaging AM equipment. 
IP violation attacks aim to illegally replicate 3D objects or the manufacturing process itself. 
Additionally, several articles discuss misuse of 3D printers for manufacturing of illegal items, e.g., firearms, or components of explosive devices; 
The analysis in a recent paper~\cite{yampolskiy2017subtractive} shows that there are no specific technical challenges of the latter (but rather legal and policy issues).
For this paper, only sabotage is considered.

To our knowledge, the first proof of concept compromise of a desktop 3D printer was presented at XCon2013\footnote{XCon2013 speakers: \href{http://xcon.xfocus.org/XCon2013/speakers.html}{http://xcon.xfocus.org/XCon2013/speakers.html}} by Xiao Zi Hang (Claud Xiao)~\cite{xiao2013security}. Among other things, the keynote argues that the size (and thus integrability) of a printed part can be modified, the temperature of the filament extruder can be manipulated, etc.

Several publications have analyzed 3D printers and 3D printing processes for vulnerabilities.
Turner et al., 2015~\cite{turner2015bad} found that networking and communication systems lack integrity checks when receiving the design files. 
Moore et al., 2016~\cite{moore2016vulnerability} identified numerous vulnerabilities in software, firmware, and communication protocol commonly used in desktop 3D printers; these can be potentially be exploited.
Do et al., 2016~\cite{do2016data} have shown that communication protocols employed by desktop 3D printers can be exploited, enabling the retrieval of current and previously printed 3D models, halting an active printing job, or submitting a new one.
Belikovetsky et al., 2017~\cite{belikovetsky2017dr0wned} used a fishing attack to install a backdoor that enabled arbitrary, targeted manipulations of design files by a remote adversary.
Sturm et al., 2014~\cite{sturm2014cyber} used malware pre-installed on a computer to automate the manipulation of STL files. 
Moore et al., 2016~\cite{moore2017implications} used malicious firmware to modify and substitute a printed 3D model.

A growing body of publications discusses how a manufactured part's quality can be compromised. 
The majority of publications focuses on Fused Deposition Modeling (FDM), commonly used in desktop 3D printers.
Sturm et al., 2014~\cite{sturm2014cyber} demonstrated that a part's tensile strength can be degraded by introducing defects such as voids (internal cavities).
Zeltmann et al., 2016~\cite{zeltmann2016manufacturing} showed that similar results can be achieved by printing part of the structure with the contaminated material.
Belikovetsky et al, 2017~\cite{belikovetsky2017dr0wned} proposed to degrade a part's fatigue life; the authors argue that the defect's size, geometry, and location are factors in the degradation. 
Yampolskiy et al, 2015~\cite{yampolskiy2015security} argued that the anisotropy characteristic of 3D printed parts can be misused to degrade a part's quality, if an object is printed in the wrong orientation.
Zeltmann et al., 2016~\cite{zeltmann2016manufacturing} have experimentally shown the impact of this attack on part's tensile strength, using 90 and 45 degree rotations of the printed model.
Chhetri et al., 2016~\cite{chhetri2016kcad} introduced a skew along one of the build axes as an attack.
Moore et al., 2016~\cite{moore2017implications} modified the amount of extruded source material to compromise the printed object's  geometry.
Pope et al., 2016~\cite{pope2016hazard} identified that indirect manipulations like the modification of network command timing and energy supply interruptions can be potential means of sabotaging a part. 
Yampolskiy et al, 2015~\cite{yampolskiy2015security} discussed various metal AM process parameters whose manipulation can sabotage a part's quality; for the powder bed fusion (PBF) process, the identified parameters include heat source energy, scanning strategy, layer thickness, source material properties like powder size and form, etc.
Yampolskiy et al., 2016~\cite{yampolskiy2016using} argued that in the case of metal AM, manipulations of manufacturing parameters can not only sabotage a part's quality, but also damage the AM machine, or lead to the contamination of its environment.

Chhetri et al, 2016~\cite{chhetri2016kcad} present the only method published so far for the detection of sabotage attacks.
The authors use the acoustic side-channel inherent to the FDM process. The reported detection rate of object modifications is 77.45\%.

{\footnotesize \bibliographystyle{acm}
\bibliography{bib}}

\begin{thebibliography}{10}

\bibitem{agrawal2007trojan}
{\sc Agrawal, D., Baktir, S., Karakoyunlu, D., Rohatgi, P., and Sunar, B.}
\newblock Trojan detection using ic fingerprinting.
\newblock In {\em Security and Privacy, 2007. SP'07. IEEE Symposium on\/}
  (2007), IEEE, pp.~296--310.

\bibitem{belikovetsky2017dr0wned}
{\sc Belikovetsky, S., Yampolskiy, M., Toh, J., and Elovici, Y.}
\newblock dr0wned-cyber-physical attack with additive manufacturing.
\newblock {\em arXiv preprint arXiv:1609.00133\/} (2016).

\bibitem{blackman20141st}
{\sc Blackman, J.}
\newblock The 1st amendment, 2nd amendment, and 3d printed guns.

\bibitem{brown2016legal}
{\sc Brown, A., Yampolskiy, M., Gatlin, J., and Andel, T.~R.}
\newblock {Legal Aspects of Protecting Intellectual Property in Additive
  Manufacturing}.
\newblock Unpublished manuscript, 2016.

\bibitem{cano2005review}
{\sc Cano, P., Batlle, E., Kalker, T., and Haitsma, J.}
\newblock A review of audio fingerprinting.
\newblock {\em Journal of VLSI signal processing systems for signal, image and
  video technology 41}, 3 (2005), 271--284.

\bibitem{chhetri2016kcad}
{\sc Chhetri, S.~R., Canedo, A., and Al~Faruque, M.~A.}
\newblock Kcad: kinetic cyber-attack detection method for cyber-physical
  additive manufacturing systems.
\newblock In {\em Proceedings of the 35th International Conference on
  Computer-Aided Design\/} (2016), ACM, p.~74.

\bibitem{depoorter2013intellectual}
{\sc Depoorter, B.}
\newblock Intellectual property infringements \& 3d printing: Decentralized
  piracy.
\newblock {\em Hastings LJ 65\/} (2013), 1483.

\bibitem{ding2004k}
{\sc Ding, C., and He, X.}
\newblock K-means clustering via principal component analysis.
\newblock In {\em Proceedings of the twenty-first international conference on
  Machine learning\/} (2004), ACM, p.~29.

\bibitem{do2016data}
{\sc Do, Q., Martini, B., and Choo, K.-K.~R.}
\newblock A data exfiltration and remote exploitation attack on consumer 3d
  printers.
\newblock {\em IEEE Transactions on Information Forensics and Security 11}, 10
  (2016), 2174--2186.

\bibitem{ey2016how}
{\sc {Ernst \& Young}}.
\newblock {How will 3D printing make your company the strongest link in the
  value chain?}
\newblock Tech. rep., 2016.

\bibitem{faruque2016acoustic}
{\sc Faruque, M.~A., Chhetri, S.~R., Canedo, A., and Wan, J.}
\newblock Acoustic side-channel attacks on additive manufacturing systems.
\newblock In {\em Proceedings of the ACM/IEEE Internation Conference on
  Cyber-Physical Systems (ICCPS' 16)\/} (2016).

\bibitem{frigo1998fftw}
{\sc Frigo, M., and Johnson, S.~G.}
\newblock Fftw: An adaptive software architecture for the fft.
\newblock In {\em Acoustics, Speech and Signal Processing, 1998. Proceedings of
  the 1998 IEEE International Conference on\/} (1998), vol.~3, IEEE,
  pp.~1381--1384.

\bibitem{holbrook2014digital}
{\sc Holbrook, T.~R., and Osborn, L.}
\newblock Digital patent infringement in an era of 3d printing.
\newblock {\em UC Davis Law Review, Forthcoming\/} (2014).

\bibitem{johnson2013print}
{\sc Johnson, J.~J.}
\newblock Print, lock, and load: 3-d printers, creation of guns, and the
  potential threat to fourth amendment rights.
\newblock {\em Browser Download This Paper\/} (2013).

\bibitem{mcmullen2014worlds}
{\sc McMullen, K.~F.}
\newblock Worlds collide when 3d printers reach the public: Modeling a digital
  gun control law after the digital millenium copyright act.
\newblock {\em Mich. St. L. Rev.\/} (2014), 187.

\bibitem{moore2016vulnerability}
{\sc Moore, S., Armstrong, P., McDonald, T., and Yampolskiy, M.}
\newblock Vulnerability analysis of desktop 3d printer software.
\newblock In {\em Resilience Week (RWS), 2016\/} (2016), IEEE, pp.~46--51.

\bibitem{moore2017implications}
{\sc Moore, S.~B., Glisson, W.~B., and Yampolskiy, M.}
\newblock Implications of malicious 3d printer firmware.
\newblock In {\em Proceedings of the 50th Hawaii International Conference on
  System Sciences\/} (2017).

\bibitem{pearson1901liii}
{\sc Pearson, K.}
\newblock Liii. on lines and planes of closest fit to systems of points in
  space.
\newblock {\em The London, Edinburgh, and Dublin Philosophical Magazine and
  Journal of Science 2}, 11 (1901), 559--572.

\bibitem{pope2016hazard}
{\sc Pope, G., and Yampolskiy, M.}
\newblock {A Hazard Analysis Technique for Additive Manufacturing}.
\newblock In {\em Better Software East Conference\/} (2016).

\bibitem{GE2015faa}
{\sc Reports, G.}
\newblock The faa cleared the first 3d printed part to fly in a commercial jet
  engine from ge.
\newblock Tech. rep., 2015.

\bibitem{sturm2014cyber}
{\sc Sturm, L., Williams, C., Camelio, J., White, J., and Parker, R.}
\newblock Cyber-physical vunerabilities in additive manufacturing systems.
\newblock {\em Context 7\/} (2014), 8.

\bibitem{turner2015bad}
{\sc Turner, H., White, J., Camelio, J.~A., Williams, C., Amos, B., and Parker,
  R.}
\newblock Bad parts: Are our manufacturing systems at risk of silent
  cyberattacks?
\newblock {\em Security \& Privacy, IEEE 13}, 3 (2015), 40--47.

\bibitem{xiao2013security}
{\sc {Xiao Zi Hang (Claud Xiao)}}.
\newblock Security attack to 3d printing, 2013.
\newblock Keynote at XCon2013.

\bibitem{yampolskiy2014intellectual}
{\sc Yampolskiy, M., Andel, T.~R., McDonald, J.~T., Glisson, W.~B., and
  Yasinsac, A.}
\newblock Intellectual property protection in additive layer manufacturing:
  Requirements for secure outsourcing.
\newblock In {\em Proceedings of the 4th Program Protection and Reverse
  Engineering Workshop\/} (2014), ACM, p.~7.

\bibitem{yampolskiy2014towards}
{\sc Yampolskiy, M., Andel, T.~R., McDonald, J.~T., Glisson, W.~B., and
  Yasinsac, A.}
\newblock {Towards Security of Additive Layer Manufacturing}, 2014.
\newblock WiP presented at The 30st Annual Computer Security Applications
  Conference (ACSAC) 2014.

\bibitem{yampolskiy2017subtractive}
{\sc Yampolskiy, M., King, W.~E., Pope, G., Belikovetsky, S., and Elovici, Y.}
\newblock {Subtractive vs. Additive Manufacturing -- Similarities and
  Differences from the Security Perspective}.
\newblock Unpublished manuscript (under review), 2017.

\bibitem{yampolskiy2015security}
{\sc Yampolskiy, M., Schutzle, L., Vaidya, U., and Yasinsac, A.}
\newblock Security challenges of additive manufacturing with metals and alloys.
\newblock In {\em Critical Infrastructure Protection IX}. Springer, 2015,
  pp.~169--183.

\bibitem{yampolskiy20163dpaaw}
{\sc Yampolskiy, M., Skjellum, A., Kretzschmar, M., Overfelt, R.~A., Sloan,
  K.~R., and Yasinsac, A.}
\newblock Using 3d printers as weapons.
\newblock {\em International Journal of Critical Infrastructure Protection\/}
  (2016).

\bibitem{yampolskiy2016using}
{\sc Yampolskiy, M., Skjellum, A., Kretzschmar, M., Overfelt, R.~A., Sloan,
  K.~R., and Yasinsac, A.}
\newblock {Using 3D Printers as Weapons}.
\newblock {\em International Journal of Critical Infrastructure Protection
  14\/} (2016), 58--71.

\bibitem{zeltmann2016manufacturing}
{\sc Zeltmann, S.~E., Gupta, N., Tsoutsos, N.~G., Maniatakos, M., Rajendran,
  J., and Karri, R.}
\newblock Manufacturing and security challenges in 3d printing.
\newblock {\em JOM\/} (2016), 1--10.

\end{thebibliography}


\end{document}